\documentclass[preprint]{aastex}

\usepackage{amsmath,amssymb,amsfonts,latexsym}

\newcommand{ \bra }{\left(}       	
\newcommand{ \ket }{\right)}      	
\newcommand{ \braa }{\left[ }     	
\newcommand{ \kett }{\right]}     	
\newcommand{\deriv}[2]{\frac{\mathrm{d} #1}{\mathrm{d} #2}} 
\newcommand{\vrel}{v_{\rm{rel}}}
\newcommand{\ud}{u_\mathrm{d}}

\newcommand{\rd}{r_\mathrm{d}}

\shorttitle{ORBITAL CIRCULARIZATION OF PLANET ACCRETING DISK GAS}

\shortauthors{Kikuchi, Higuchi, Ida}

\begin{document}

\title{
ORBITAL CIRCULARIZATION OF A PLANET ACCRETING DISK GAS:
FORMATION OF DISTANT JUPITERS IN CIRCULAR ORBITS BASED ON CORE ACCRETION MODEL
}
\author{AKIHIRO KIKUCHI\altaffilmark{1}, ARIKA HIGUCHI\altaffilmark{1}, AND SHIGERU IDA\altaffilmark{2}}
\affil{\altaffilmark{1}Department of Earth and Planetary Sciences, Tokyo Institute of Technology, Meguro-ku, Tokyo 152-8551, Japan}
\affil{\altaffilmark{2}Earth-Life Science Institute, Tokyo Institute of Technology, Meguro-ku, Tokyo 152-8550, Japan}
\email{kikuchi.a@geo.titech.ac.jp, higuchia@geo.titech.ac.jp, ida@elsi.jp}

\begin{abstract}
Recently, gas giant planets in nearly circular orbits
with large semimajor axes ($a \sim 30$--1000AU) have been detected
by direct imaging.
We have investigated orbital evolution in a formation scenario
for such planets, based on core accretion model:
i) Icy cores accrete from planetesimals at $\la 30$AU,
ii) they are scattered outward by an emerging nearby gas giant
to acquire highly eccentric orbits, and iii) their orbits are circularized
through accretion of disk gas in outer regions,
where they spend most of time.
We analytically derived equations to describe the orbital
circularization through the gas accretion.
Numerical integrations of these equations show that
the eccentricity decreases by a factor of more than 5
during the planetary mass increases by a factor of 10.
Because runaway gas accretion increases planetary mass
by $\sim 10$--300, the orbits are sufficiently circularized.
On the other hand, $a$ is reduced
at most only by a factor of 2, leaving the planets in outer regions.
If the relative velocity damping by shock is considered,
the circularization is slowed down, but still efficient enough.
Therefore, this scenario potentially accounts for
the formation of observed distant jupiters in nearly circular orbits.
If the apocenter distances of the scattered cores
are larger than the disk sizes,
their $a$ shrink to a quarter of the disk sizes;
the $a$-distribution of distant giants could reflect
outer edges of the disks in a similar way that
those of hot jupiters may reflect inner edges. 
\end{abstract}


\section{Introduction}
Distant extrasolar gaseous giant planets in nearly circular orbits have been detected by direct imaging observations 
in several systems (e.g.,\citealp{kalas08,marois08,Kuzuhara13}). 
In the conventional core accretion model, it is difficult to form cores that are massive enough to undergo runaway gas accretion at $\ga 30$AU within disk lifetime 
($\sim$ a few million years)\footnote{Recently, "pebble accretion" has been proposed, which is a rapid growth process accreting small bodies suffering strong gas drag \citep[e.g.,][and references therein]{johansen12,youdin13}. 
If pebble accretion works well in outer disk regions, 
cores could be formed even outside of 30 AU.
This possibility should also be pursued, although it is not discussed here.}, 
because
the core growth timescale is roughly proportional to a cube of the distance from the central star (e.g., \citealp{ida04a}).
Although the cores cannot be formed in such distant regions,
gas giant planets formed interior to 30AU can be scattered by
other giant planets to 
attain semimajor axes $a \ga 100$AU (e.g., \citealp{marzari00,nagasawa08}).
Because the dynamical energy is lower in outer regions,
$a$ of scattered planets are broadly distributed up to $\sim 1000$AU.
However, due to the conservation of the total angular momentum,
the eccentricities $e$ of the scattered orbits must be excited to be close to unity.
While disk-planet interactions tend to damp such high values of $e$,
they may not be efficient enough to account for the observed low eccentricities 
of distant, gas giant planets, because local protoplanetary disk mass
may not be massive enough in the distant regions
\citep{ida13}.

This difficulty has raised the possibility of formation of gas giants by
disk gravitational instability (e.g., \citealp{boss01,helled14} and references therein).
\citet{kratter10} showed that if disk instability forms planetary mass clumps, it would form more abundant brown-dwarfs and M-star companions.
A population synthesis simulation based on the disk instability model (\citealp{forgan13})
showed that most of such brown-dwarfs and M-star companions may be retained in 
outer regions.
However, it is not consistent with direct imaging surveys so far done.
Furthermore, the observationally clear correlation between fraction of stars with gas giants
and stellar metallicity (\citealp{fischer05}) is not easy for the
disk instability model to explain, while the correlation is
consistent with the core accretion model (e.g., \citealp{ida04b}).

Based on the core accretion model, \citet{crida09} 
proposed outward type II migration 
for the origin of the distant gas giants in nearly circular orbits.
However, the outward type II migration requires a pair of giant planets 
in a common gap with the
inner one more massive than the outer one and appropriate disk conditions.

\citet{ida13} found out another path
to form the distant gas giants in nearly circular orbits, 
based on the core accretion model:
outward scattering of cores by a nearby gas giant
followed by accretion of gas in outer regions.
Because the orbital circularization through gas accretion was
also shown by a hybrid N-body and 2D hydrodynamical simulation
(E. Thommes 2010, private communication),
this path is one of promising mechanisms.
Details of the path they found are as follows.
Oligarchic growth produces
similar-sized multiple cores (\citealp{kokubo98}).
Once some core starts runaway gas accretion,
the planet's mass rapidly increases \citep[e.g.,][]{bodenheimer86,ikoma00}.
When the mass increase is fast enough, the planet
undergoes close encounters with nearby cores 
to strongly scatter them rather than shepherd them
(\citealp{zhou07,shiraishi08}).
Some cores
are scattered to large distance where the surface densities of both residual
planetesimals and gas are relatively low.  
If the scattered planet is a gas giant, its mass could be
comparable to or larger than the local disk mass,
because the scattering may occur in a late disk 
evolution stage after the formation of the gas giant. 
Then, eccentricity damping due to dynamical friction 
from local disk gas is inefficient.

On the other hand, in the core scattering model, 
cores' masses are usually well below the local disk mass.
However, since the scattered cores have highly eccentric orbits
and the relative velocity between the cores and the disk gas should be highly
supersonic, dynamical friction from the local disk gas 
is less efficient (\citealp{ostriker99,papaloizou00,muto11}).
\citet{muto11} showed that
the dynamical friction timescale in supersonic regime for a planet with mass $M_p$,
orbital eccentricity $e$ and semimajor axis $a$ in a gas disk with surface density $\Sigma$ is
\begin{equation}
\tau_{\rm DF} \sim \frac{1}{8\pi}\left(\frac{M_\ast}{\Sigma a^2}\right)
\left(\frac{M_\ast}{M_p}\right) \left(\frac{c_s}{v_{\rm K}}\right) 
\left(\frac{v}{v_{\rm K}}\right)^{3}T_{\rm K}
\sim 
10^4 \left(\frac{\Sigma a^2}{0.01M_\ast}\right)^{-1}
\left(\frac{M_p}{10 M_{\oplus}}\right)^{-1} \left(\frac{c_s/v_{\rm K}}{0.1}\right) 
e^3 T_{\rm K},
\end{equation}
where $M_\ast$ is the host star's mass, $c_s$ is local sound velocity, and
$v_{\rm K}$ and $T_{\rm K}$ are Kepler velocity and its orbital period at $a$. 
For a planet with mass $10M_{\oplus}$ in a highly eccentric orbit ($e \sim 1$) with $a \sim 100$AU,
$\tau_{\rm DF}$ is as long as $\sim 10^7$ years, which is longer than 
an observationally inferred disk lifetime $\sim {\rm a \; few} \times 10^6$ years.
Thereby, the dynamical friction from local disk gas may be less
effective than the orbital circularization via planetary gas accretion, 
which we discuss in this paper (see section \ref{sec:pop}).
Furthermore, since the dynamical friction also damps semimajor axis
efficiently in the course of eccentricity damping from values close to unity,
we may not be able to retain the cores in outer regions.

Reduction in the planetesimal
accretion rate decreases the critical core mass for 
the onset of gas accretion (e.g., \citealp{ikoma00}).
The scattered cores in highly
eccentric orbits spend most of time at large distances, 
where planetesimal accretion rate is significantly low,
so that it is possible for the scattering to trigger gas accretion onto the cores. 
In the case of highly eccentric orbits, since the cores
spend most of time near apocenters at large distance, 
their orbits are circularized there through accretion
of local gas with higher specific angular momentum.
As a result, it is expected that 
the cores' orbits are circularized keeping 
their apocenters almost fixed, in the course of gas accretion. 
We will show that the apocenter shrinks 
because the semimajor axis slightly shrinks due to
energy dissipation by collision between the disk gas and the planet
and that the energy dissipation accelerates the orbital circularization
(in the case of moderate eccentricity, the energy dissipation
is more important for the orbital circularization
than accretion of high angular momentum gas).

Assuming that the orbits of the scattered cores are quickly
circularized to the degree that depends on the ratio between the planet mass
and local disk mass with their semimajor axes kept fixed,
\citet{ida13} performed a population synthesis simulation
to predict statistical distributions of distant gas giants formed by this mechanism. 
They showed that the fraction of systems with the distant gas giants
is a $\sim 0.1$--1\% and most of them have low eccentricities ($e \la 0.1$).
Although the fraction is further lower, systems with multiple distant
gas giants are also formed, because a single gas giant can scatter
multiple cores in inner regions.
HR8799 system has four distant gas giants and 
the outer three planets could be in 4:2:1 resonance.
In the scattering core model, formation of four distant gas giants  
is extremely rare and probability for capture into the resonances during
inward migration associated with eccentricity damping is not clear.
However,  the inner two planets have semimajor axes 
$\sim$ 15 and 27 AU, which could be formed {\it in situ} without
the scattering process.
Formation of HR8799 system by core accretion scenario is
a very interesting problem that should be addressed in the future.  
In this paper, we focus on a fundamental process of
orbital circularization of an isolated planet through gas accretion. 

In the core scattering model, 
positive correlation between the
semimajor axis and the mass of the distant gas giants is predicted.
The critical planet mass for gap opening is higher
in larger orbital radius (e.g., \citealp{ida04a}).
If the gap opening halts growth of gas giants, 
the correlation is established, as shown in the population
synthesis calculation in \citet{ida13}.

Note that \citet{ida13} assumed that the eccentricities of the scattered cores
are efficiently damped without any decrease in semimajor axes,
implicitly assuming very efficient damping due to gas accretion,
although they did not incorporate detailed orbital evolution
by the gas accretion. 
As we will show, the eccentricity damping due to gas accretion
is indeed efficient, 
while the degree of damping depends on how much the planets
grow by accreting gas and semimajor axes are also damped
by a factor of $\sim 2$.
The population synthesis simulation must be
improved by incorporating the damping formulas due to gas accretion
derived in this paper, in order to discuss the distribution of
distant gas giants in comparison with
observation when the number of discovered planets
becomes large enough for statistical arguments.

Here, we investigate the orbital circularization of the scattered
planets during gas accretion through detailed analytical calculations.
Section 2 describes the assumptions of gas accretion onto the cores.
In section 3, we analytically derive the formulas for the orbital evolution 
in the course of gas accretion.
In section 4, we describe the orbital evolution by numerically solving the formulas.
In section 5, we show some results of the population synthesis
calculations, by incorporating the prescriptions of orbital
evolution through the mass growth due to accretion of disk gas.
Section 6 is devoted for summary.

\section{Model}

We start our calculation from the stage at which
a core has already been scattered outward by a gas giant to attain
eccentricity close to unity ($e_{\rm ini}$) 
and semimajor axis ($a_{\rm ini}$) that is much larger than 
the original one ($a_{\rm ori}$).
Note that the pericenter distance
of the scattered planet's orbit must be close to $a_{\rm ori}$:
$a_{\rm ori} \simeq q_{\rm ini} = a_{\rm ini}(1-e_{\rm ini})$.
Since $a_{\rm ori}$ should be close to the gas giant's orbital radius,
we can regards that $q_{\rm ini} \sim 1$--10AU, based on core accretion
model (e.g., \citealp{ida04a}).
As we will show later, the pericenter distance of the scattered
planet quickly increases.
Accordingly, the planet immediately 
becomes isolated from the perturbing gas giant, so that
we neglect its further perturbations.
We do not calculate the initial scattering process by the gas giant,
but study evolution of $e$ and $a$ of the scattered planet
due to accretion of gas,
for given $e_{\rm ini}$, $a_{\rm ini}$ and disk radius $\rd$,
to derive general formulas for the orbital evolution.
In the following, we explain our prescriptions for
accretion of gas onto the scattered planet. 

After a core mass exceeds a critical core mass, pressure gradient no longer
supports gas envelope of the planet against the planetary gravitational force
and quasi-static contraction of the gas envelope starts (e.g., \citealp{mizuno80,
bodenheimer86}). 
The critical core mass is given by (\citealp{ikoma00})
\begin{equation}
M_{\rm c,crit} \simeq 10 
\left( \frac{\dot{M}_{\rm c}}{M_\oplus /10^6 \mbox{yr}} \right)^{(0.2-0.3)} 
\left( \frac{\kappa}{\kappa_{0}} \right)^{(0.2-0.3)} 
M_{\oplus},
\label{eq:M_c_crit}
\end{equation}
where $\kappa$ is opacity of the gas envelope and
$\kappa_{0}$ is that of the minimum-mass solar nebula model (\citealp{hayashi81}).
Since the planetesimal accretion rate, $\dot{M}_{\rm c}$, determines
heat energy source to support the envelope,
a lower value of $\dot{M}_{\rm c}$ leads to a smaller value of $M_{\rm c,crit}$.
In general, planetesimal accretion rate rapidly decreases with orbital radius
(e.g., \citealp{ida04a}).
After a core is scattered outward to acquire high orbital eccentricity,
the core spends most of time at much larger orbital radii than 
near the original location. 
Thereby, an orbit-averaged value of $\dot{M}_{\rm c}$ is 
significantly lowered and it is likely that quasi-static contraction of 
gas envelope is initiated by the outward scattering.

According to the quasi-static contraction, disk gas can be supplied to Hill radius or
Bondi radius of the planet.
This means that gas accretion rate onto the planet
is regulated by heat transfer through the envelope
rather than environmental disk gas conditions except in the final stage in which
the contraction is very fast.
The (Kelvin-Helmholtz) timescale of the envelope contraction is given by (\citealp{ikoma00,ikoma06})
\begin{equation}
\tau_{\rm KH} \simeq 10^{10} \left( \frac{M}{M_\oplus} \right)^{-(3-4)}
\left(\frac{\kappa}{\kappa_{\rm ini}}\right){\rm years}.
\label{eq:KHtime}
\end{equation}
From these arguments, we assume that gas accretion rate does not
depend on the position of the eccentric orbit during an orbital period, 
although environmental disk conditions
considerably change during one orbital cycle for highly eccentric orbits. 

In general, the dependence of $\tau_{\rm KH}$ on orbital radius $r$ is weak
for radiation-dominated envelope \citep[e.g.][]{ikoma00}.
If convective envelope develops, envelope contraction rate
can be affected by disk temperature and density \citep{ikoma01, piso14}.
However, even if $\tau_{\rm KH}$ has the $r$-dependence,  
the dependence may be smoothed out when $\tau_{\rm KH}$ is longer than 
the orbital period, that is, when $M \la 100M_{\oplus}$, 
because the response time of the envelope structure is 
given by $\tau_{\rm KH}$.
As we show in the following, gas accretion rate onto the planet
may be regulated by disk gas supply rather than by envelope contraction
for $M \ga 100M_{\oplus}$.

Note that in the highly eccentric orbit, Bondi and Hill radii significantly
change during one orbital circulation.  
The change might also induce oscillation of gas envelope that could
affect gas accretion and heat generation/cooling.
Investigation of this effect is left for a future work.
We will only assume the constant accretion rate during one orbit inside the disk,
but not adopt any particular form of $\tau_{\rm KH}$.

When the envelope contraction is faster than the supply
of gas and the supply has the $r$-dependence,
the assumption of the constant accretion rate is violated.
The supply can be limited by global disk accretion and
Bondi accretion.
The limit by global disk accretion becomes important
for $M \ga 100M_{\oplus}$, because
the quasi-static contraction rate is given by
$\dot{M}_{\rm KH} \sim M/\tau_{\rm KH} \sim 10^{-10}(M/M_{\oplus})^{(4-5)}(\kappa/\kappa_{\rm ini})^{-1}M_{\oplus}/{\rm yr}$ (Eq.~(\ref{eq:KHtime}))
and 
the observationally inferred typical value of
disk accretion rates onto T Tauri stars
is $\dot{M}_{\rm disk} \sim 10^{-8}M_{\odot}/{\rm yr} \sim 3 \times 10^{-3}M_{\oplus}/{\rm yr}$.
However, the $r$-dependence of disk accretion rate 
is very weak in the regions of $r \ll r_{\rm d}$, where
a steady-accretion-disk approximation is valid.  

The Bondi accretion rate is given by
$\dot{M}_{\rm Bondi} \sim \pi \rho_g (GM/v^2)^2 v$
where $\rho_g$ is disk gas density
and $v$ is the relative velocity between the planet and disk gas.
Both $\rho_g$ and $v$ sensitively depend on $r$.
In general, $\dot{M}_{\rm Bondi} < \dot{M}_{\rm KH}$ for high $e$
and large $M_{\rm p}$.
But, our calculations start from small $M_{\rm p}$ and
$e$ is already damped when $M_{\rm p}$ becomes large.
We found that in most of orbital evolution we consider,
$\dot{M}_{\rm Bondi} \ga \dot{M}_{\rm KH}$
and the supply limit by Bondi accretion does not occur.

Thus, our assumption of constant gas accretion may be justified.
As a result of the time-independent gas accretion rate, 
the planet accretes gas preferentially in outer regions
where the planet spends most of time.
The specific angular momentum of the planetary orbit is given by
$\ell_{\rm p} = \sqrt{GM_*a(1-e^2)} \sim \sqrt{2GM_*q}$,
where $q=a(1-e)$ is the pericenter distance and $e\sim 1$ is assumed.
That of the local gas near the apocenter ($Q = a(1+e)\simeq 2a$) 
is given by $\ell_{\rm g} \sim \sqrt{2GM_*a}$ 
(we assume circular Keplerian motion for the disk gas).
Since $q = a(1-e) \ll a$ for $e\sim 1$, 
the planet's specific angular momentum is increased 
by the accretion of local gas.
Then, the planetary orbit tends to be circularized
with the apocenter distance ($Q$) fixed.

If the orbit deviates from the disk with a finite size $r_{\rm d}$,
beyond which gas density is significantly declined, 
we halt gas accretion at $r > r_{\rm d}$.
In the following derivations, we consider two cases: 
i) the apocenter is inside the disk ($Q <\rd$)
and ii) it is outside the disk ($Q >\rd$).
We will refer to cases i) and ii) as "embedded case" and 
"deviated case," respectively.
In the "deviated case,"
the planet mostly accretes gas
 at $r \sim r_{\rm d}$ and planetary orbits
 tend to be fitted to circular orbits at $r_{\rm d}$
 rather than those at $Q$.

The specific orbital energies of the planet and local gas
near the apocenter are 
$\epsilon_{\rm p} = - GM_*/2a$ and 
$\epsilon_{\rm g} = - GM_*/2a(1+e) \sim  - GM_*/4a$, respectively.
Near the apocenter, the planet's specific orbital energy is increased 
by the accretion of local gas near the apocenter.
However, the accretion of lower specific energy near the
pericenter is significant due to a deep potential near the pericenter,
in spite of fast passage of the pericenter. 
As we show in section \ref{sec:embedded}, in the embedded case, the orbit-averaged 
specific orbital energy of accreting gas is exactly the same as
that of the planet, irrespective of orbital eccentricity.
The planet's specific orbital energy
actually decreases if collisional dissipation between the planet and disk gas
is taken into account.
It also contributes to eccentricity damping, with
slight decay of the semimajor axis.

In embedded case, gas accretion rate onto the planet is independent of
phase of the orbits.
In the deviated case, we assume a constant gas accretion rate
at $r < r_{\rm d}$ and zero accretion rate at $r > r_{\rm d}$.
We do not assume even the value of the constant accretion rate,
because we will derive orbital evolution as a function of planetary mass $M_{\rm p}$
but not as a function of time.

The relative velocity between the planet and the disk gas would be supersonic almost everywhere 
for highly eccentric orbits with $e \ga h/r \sim 0.1$ where $h$ is the disk scale height.
For incident supersonic gas flow to stay in Hill radius or Bondi radius,
we need some energy dissipation.
Bow shock in front of the planet may provide the energy dissipation.
We will leave full hydrodynamic simulations on the bow shock for
future work and 
assume that the planet accretes disk gas in unperturbed flow
and the accretion rate is independent of orbital phase
in most of calculations.
Note, however, that the relative velocity between the gas flow
and the planet is smaller in the post-shock flow than 
in the unperturbed flow, which may make the eccentricity 
damping less efficient.
In section 4.3, we perform calculations taking into account
the effect of the shock with a simple 1D model and show that
the eccentricity damping is indeed slowed down but does not significantly
change our conclusion.

When cores are scattered by a gas giant, in early stage,
eccentricities 
are preferentially pumped up compared with inclinations.
But, if the core undergoes repeated close encounters
before its orbit is circularized, orbital inclinations are also excited.
We also calculated $e$ and $a$ evolution with non-zero inclinations.
We have found that the final values of $e$ and $a$ 
change by less than 5\% if the inclination is smaller than 30 degrees.
So, we here show the results with zero inclinations.

In summary, the assumptions 
we use in most of runs are:
\begin{enumerate}
\item
The gas disk is in Keplerian rotation.
Because the relative velocity between the gas and
the planet is generally supersonic, the gas
is hardly perturbed by the planetary gravitational perturbations.

\item
The motions of the planet and the gas disk are coplanar.
\item
The gas accretion rate onto the planet
is constant of during one orbit, so that
orbit-averaging can be done.
The planet captures the local unperturbed disk gas, conserving mass and 
angular momentum (energy is not conserved).
\item
If the gas disk has the finite size,
we truncate gas accretion during the period in which the  planet goes out of the disk.
\end{enumerate}
In section 3 through section 4.2, we adopt the above assumptions and
derive analytical formulas to describe the orbital circularization process 
through the gas accretion.
Even if we adopt assumption 4., analytical formulas are derived,
since the constant gas accretion rate is still applied at $r < \rd$ and
analytical orbit averaging can be done.
If we include the effect of shock dissipation, analytical integration
is not possible, so that we show the orbital evolution
obtained by numerical integration (section 4.3).

\section{Derivation of formulas for orbital changes}
With the assumptions 1 to 4 described in the above,
we analytically derive formulas to calculate the orbital evolution
in the form of differential equations.
Numerically integrating the differential equations,
we will show the evolution paths of $e$ and $a$ that are 
uniquely determined by their initial values and $\rd$.

According to discussions in section 2, we first calculate
changes in the angular momentum and energy of 
the planet, $\Delta L$ and $\Delta E$, during one orbital period, 
assuming that the mass accretion rate is constant with time during one orbital period.
We also assume that the changes in orbital elements are 
small enough 
over one orbital period, in other words, the mass of the captured gas 
during one orbit ($\Delta M$) is much smaller than
the instantaneous planetary mass ($M$).

The changes $\Delta L$ and $\Delta E$ are then given by
\begin{align}
\Delta L 
& \simeq \int l_{\rm gas} \; {\rm d} M
 = \Delta M \; \frac{1}{t_{\rm d}} \int_{-t_{\rm d}/2}^{t_{\rm d}/2} l_{\rm gas}  {\rm d} t, \\
\Delta E 
& \simeq \int \bra \epsilon_{\rm gas}  - \epsilon_{\rm coll} \ket {\rm d} M 
 = \Delta M \; \frac{1}{t_{\rm d}} \int_{-t_{\rm d}/2}^{t_{\rm d}/2}  \bra \epsilon_{\rm gas}  - \epsilon_{\rm coll} \ket {\rm d} t, 
\end{align}
where the integral is during one orbit, $t_{\rm d}$ is a duration 
at $r < r_{\rm d}$ ($t_{\rm d} \le T_{\rm K}$),
$t = 0$ is a pericenter passage, 
$l_{\rm gas}$ and $\epsilon_{\rm gas}$ are
specific angular momentum and orbital energy of accreting gas
and $\epsilon_{\rm coll}$ is energy dissipation by collision 
between the planet and accreting gas.

Through $\Delta L$ and $\Delta E$ during mass growth of $\Delta M$,
specific angular momentum and orbital energy of the planet, 
$\ell_{\rm p}$ and $\epsilon_{\rm p}$, are changed.
Since $\Delta L = (\Delta \ell_{\rm p} + \ell_{\rm p})(M+\Delta M) - M\ell_{\rm p} \simeq \Delta \ell_{\rm p} \cdot M + \ell_{\rm p} \Delta M$ and $\Delta E \simeq \Delta \epsilon_{\rm p} \cdot M + \epsilon_{\rm p} \Delta M$, the change rate of $\ell_{\rm p}$ and $\epsilon_{\rm p}$ of the planet in one orbital period are expressed as
\begin{align}
\frac{\Delta \ell_{\rm p}}{\ell_{\rm p}}
& \simeq \frac{\Delta L}{M\ell_{\rm p}} - \frac{\Delta M}{M} 
= \frac{\Delta M}{M} \; f_\ell, \\
\frac{\Delta \epsilon_{\rm p}}{\epsilon_{\rm p}}
& \simeq \frac{\Delta E}{M\epsilon_{\rm p}} - \frac{\Delta M}{M} 
= \frac{\Delta M}{M} \; f_\epsilon, 
\end{align}
where 
\begin{align}
f_\ell
&=\frac{1}{t_{\rm d}} \int_{-t_{\rm d}/2}^{t_{\rm d}/2} \left( \frac{l_{\rm gas}}{l_{\rm p}} -1 \right) {\rm d} t,
\label{eq:f_ell} \\
f_\epsilon
&=\frac{1}{t_{\rm d}} \int_{-t_{\rm d}/2}^{t_{\rm d}/2} \bra \frac{\epsilon_{\rm gas}}{\epsilon_{\rm p}}  - \frac{\epsilon_{\rm coll}}{\epsilon_{\rm p}} -1 \ket {\rm d} t.
\label{eq:f_epsilon}
\end{align}
Since $\ell_{\rm p} = \sqrt{GM_\ast a(1-e^2)}$ and $\epsilon_{\rm p} = -GM_\ast /2a$,
where $M_\ast$ is the host star's mass, $G$ is the gravitational constant, 
the changes $\Delta \ell_{\rm p}$ and $\Delta \epsilon_{\rm p}$ are related with 
the changes of the eccentricity and semimajor axis ($\Delta e$ and $\Delta a$) as
\begin{align}
\frac{\Delta \ell_{\rm p}}{\ell_{\rm p}}
&= \frac{\Delta(\sqrt{a})}{\sqrt{a}} + \frac{\Delta(\sqrt{1-e^2})}{\sqrt{1-e^2}}
\simeq \frac{\Delta a}{2a} - \frac{e}{1-e^2}\Delta e, \\
\frac{\Delta \epsilon_{\rm p}}{\epsilon_{\rm p}}
&= \frac{\Delta (a^{-1})}{a^{-1}}
\simeq - \frac{\Delta a}{a}.
\end{align}
So, $\Delta e$ and $\Delta a$ are given by
\begin{align}
\Delta e 
&\simeq - \frac{1-e^2}{e} \bra \frac{\Delta \ell_{\rm p}}{\ell_{\rm p}} + \frac{1}{2}\frac{\Delta \epsilon_{\rm p}}{\epsilon_{\rm p}} \ket
= - \frac{\Delta M}{M} f_e, \label{eq:Delta_e}\\
\frac{\Delta a}{a}
&\simeq - \frac{\Delta \epsilon_{\rm p}}{\epsilon_{\rm p}} 
= - \frac{\Delta M}{M} f_a, \label{eq:Delta_a}
\end{align}
where
\begin{align}
f_e
&= \frac{1-e^2}{e} \bra f_\ell + \frac{1}{2} f_\epsilon \ket,
\label{eq:f_e} \\
f_a
&= f_\epsilon.
\label{eq:f_a} 
\end{align}

We finally derive the differential equations for the orbit-averaged evolution of $e$ and $a$
in terms of planetary mass $M$:
\begin{align}
\deriv{e}{\log M}
&\simeq \frac{\Delta e}{\Delta M/M} = - f_e, 
\label{eq:A_e}\\
\deriv{\log a}{\log M}
&\simeq \frac{\Delta a/a}{\Delta M/M}= - f_a. 
\label{eq:A_a}
\end{align}
From these equations, we also obtain
\begin{equation}
\deriv{\log a}{e}\simeq \frac{f_a}{f_e}. \label{eq:A_ea}
\end{equation}

So far, we have not assumed any specific forms for $\ell_{\rm gas}$ and $\epsilon_{\rm gas}$.
Here we assume that the planet captures 
gas in circular Keplerian motion to analytically derive formulas $f_e$ and $f_a$.
(In section 4.3, we calculate $f_e$ and $f_a$ for post-shocked gas flow
using a simple 1D model.)
Note that analytical formulas of $f_e$ and $f_a$ are derived even for
deviated case where $Q > \rd$.

For unperturbed gas flow (circular Keplerian flow),
\begin{align}
\ell_{\rm gas} & = \sqrt{GM_\ast r}, \\
\epsilon_{\rm gas} & = - \frac{GM_\ast}{2r}, \\
\epsilon_{\rm coll} & = \frac{\vrel(r)^2}{2},
\end{align}
where $\vrel(r)$ is the relative velocity between the planet and the local gas, 
and $r$ is the instantaneous distance of the planet from the central star. 
The radial and tangential components of instantaneous velocity
of an eccentric Keplerian orbit of the planet at $r$ are given by
\begin{align}
v_r & = v_{\rm K} \sqrt{2-\frac{r}{a}-\frac{a}{r}(1-e^2)}, \label{eq:v_r}\\ 
v_\phi & = v_{\rm K} \sqrt{\frac{a}{r}(1-e^2)}, \label{eq:v_phi}
\end{align}
where $a$ and $e$ are the planet's semimajor axis and eccentricity.
Because the local Keplerian velocity is 
given by $v_{\rm K}=\sqrt{GM_\ast/r}$,
the square of relative velocity is
\begin{align}
\vrel(r)^2 &= v_r^2 + (v_\phi - v_{\rm K})^2 \\
&= \frac{GM_\ast}{r} \bra 3 - \frac{r}{a} - 2\sqrt{\frac{a}{r} (1-e^2) } \ket.
\label{eq:vrel} 
\end{align}

For integrating Eqs.~(\ref{eq:f_ell}) and (\ref{eq:f_epsilon}),
we convert time to eccentric anomaly using the Kepler equation.
The time average of powers of $r^{\alpha}(\alpha=1/2,-1,-3/2)$ are analytically integrated,
using the conversion:
\begin{align}
\frac{1}{t_{\rm d}} \int_{-t_{\rm d}/2}^{t_{\rm d}/2} \bra \frac{r}{a} \ket^{1/2} {\rm d}t 
& =\frac{2\sqrt{1+e}}{3} f_{1/2}(e,\ud); \nonumber \\
 f_{1/2}(e,\ud) & =\frac{4(E(k)-E(y,k)) - (1-e)(K(k)-F(y,k)) - e\sin2y \sqrt{1-k^2\sin^2y} }{\ud-e\sin \ud},   \label{eq:time_average_1/2} \\
\frac{1}{t_{\rm d}} \int_{-t_{\rm d}/2}^{t_{\rm d}/2} \bra \frac{r}{a} \ket^{-1} {\rm d}t
& =f_{-1}(e,\ud); \; f_{-1}(e,\ud)=\frac{\ud}{\ud-e\sin \ud}, 
\label{eq:time_average_-1} \\
\frac{1}{t_{\rm d}} \int_{-t_{\rm d}/2}^{t_{\rm d}/2} \bra \frac{r}{a} \ket^{-3/2} {\rm d}t
& =\frac{2}{\sqrt{1+e}}f_{-3/2}(e,\ud); \; f_{-3/2}(e,\ud)=\frac{K(k)-F(y,k)}{\ud-e\sin \ud},
\label{eq:time_average_-3/2} 
\end{align}
where $k \equiv \sqrt{2e/(1+e)}$, $y \equiv (\pi-\ud)/2$,
$K(k)$ is the complete elliptic integral of the first kind, $E(k)$ is the complete elliptic integral of the second kind, $F(y,k)$ is the elliptic integral of the first kind, 
$E(y,k)$ is the elliptic integral of the second kind, and
$\ud$ is the maximum eccentric anomaly ($0<\ud<\pi$)
within the disk $(r < \rd)$, which is given by
\begin{eqnarray}
\ud \equiv \left\{ \begin{array}{ll}
\cos^{-1} \braa \frac{1}{e} \bra 1- \frac{\rd}{a} \ket \kett
& [{\rm for} \hspace*{8pt} Q>\rd], \\
\pi
& [{\rm for} \hspace*{8pt} Q <\rd ]. \\
\end{array} \right.
\end{eqnarray}
In embedded case, $\ud=\pi$ and $t_{\rm d} = T_{\rm K}$,
while $\ud < \pi$ and $t_{\rm d} < T_{\rm K}$, depending on $r_{\rm d}$,
in deviated cse.

With Eqs.~(\ref{eq:time_average_1/2}) to (\ref{eq:time_average_-3/2}),
Eqs.~(\ref{eq:f_ell}) and (\ref{eq:f_epsilon}) are written as
\begin{align}
f_\ell(e,\ud)
&=\frac{2}{3\sqrt{1-e}}f_{1/2}(e,\ud) -1, \\
f_\epsilon(e,\ud)
&=4 f_{-1}(e,\ud) - 4\sqrt{1-e}f_{-3/2}(e,\ud) -2.
\end{align}
Note that $f_\ell$ and $f_\epsilon$ are functions of only $e$,
independent of $a$, in embedded case.
Even in deviated case, the $a$-dependence enters 
$f_\ell$ and $f_\epsilon$ only through the scaled quantity $a/r_{\rm d}$ in $u_{\rm d}$.

From Eqs.~(\ref{eq:f_e}) and (\ref{eq:f_a}),
\begin{align}
\deriv{e}{\log M} & = - f_e(e,\ud)
= - \frac{1-e^2}{e} \bra f_\ell(e,\ud) + \frac{1}{2} f_\epsilon(e,\ud) \ket, \label{eq:e_evol}\\
\deriv{\log a}{\log M} & = - f_a(e,\ud)
= - f_\epsilon(e,\ud). \label{eq:a_evol}
\end{align}
These equations show that while
the semimajor axis is damped only by $\Delta \epsilon_{\rm p}$,
the orbital eccentricity is damped by both $\Delta \ell_{\rm p}$ and $\Delta \epsilon_{\rm p}$.

\begin{figure}
\includegraphics[scale=0.7]{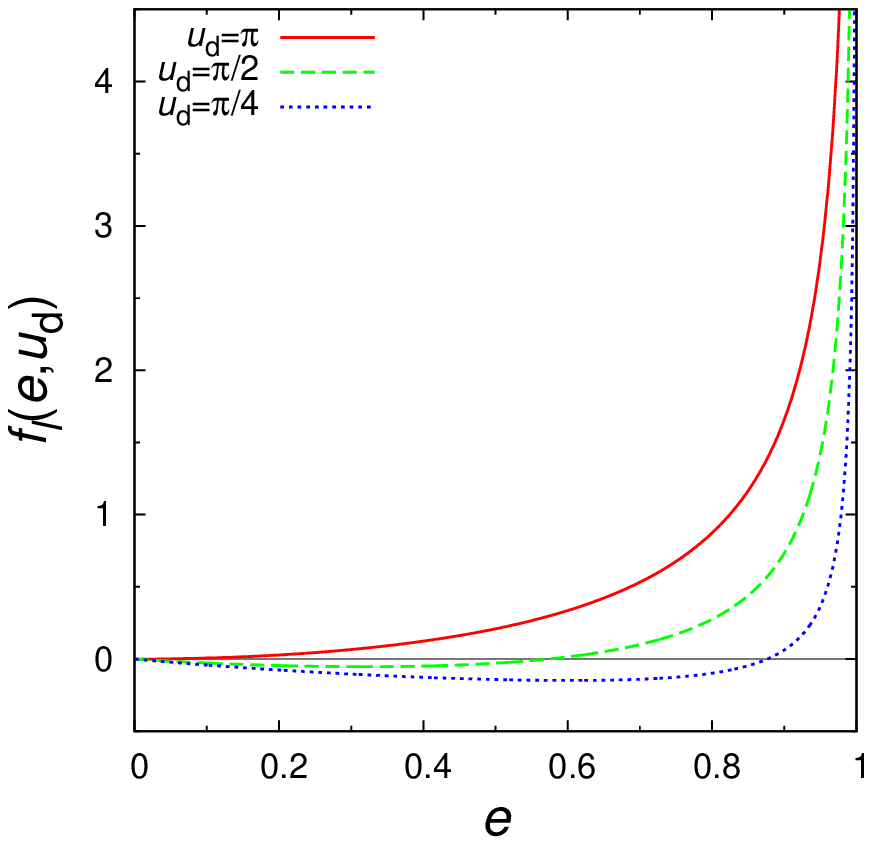}
\includegraphics[scale=0.7]{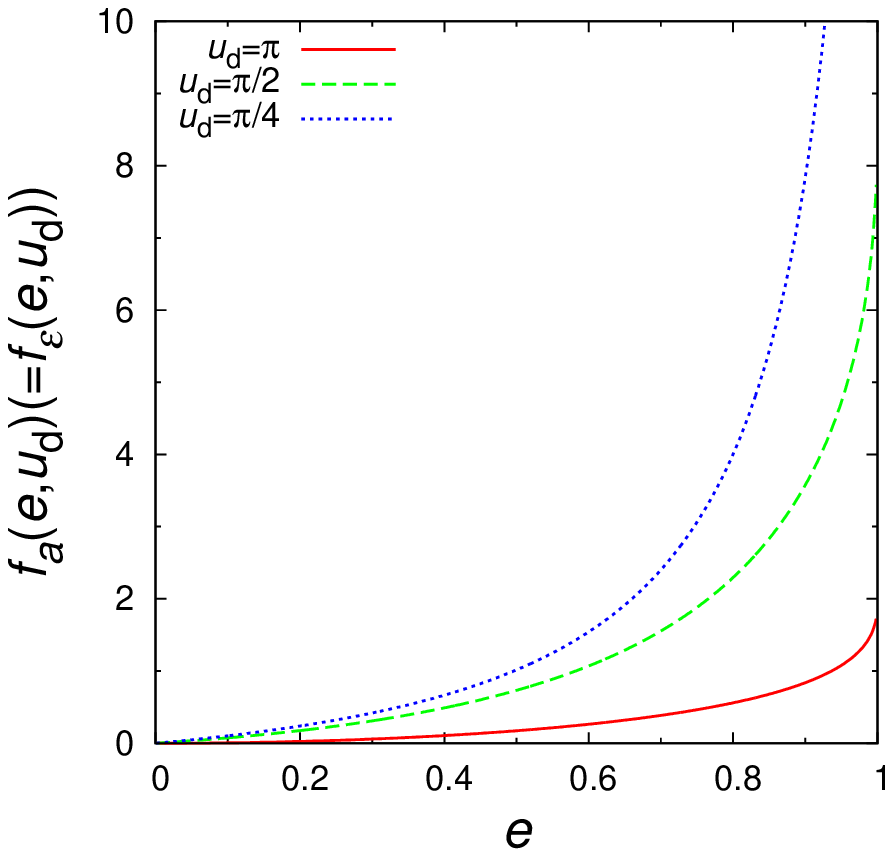}
\includegraphics[scale=0.7]{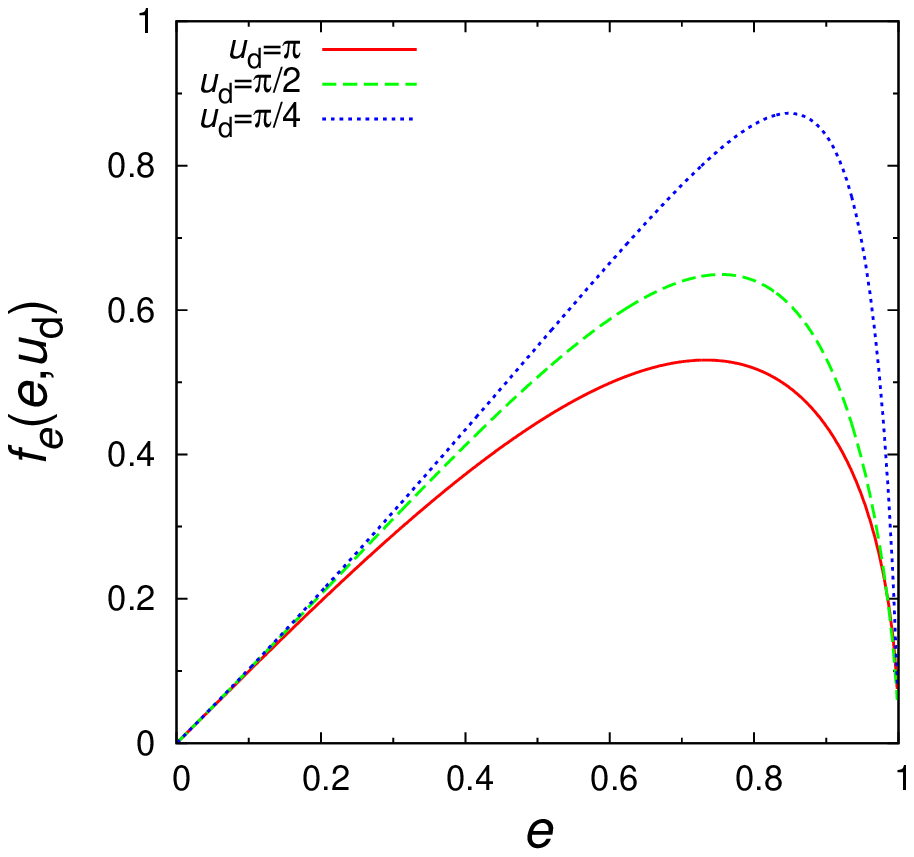}
\caption{The functions $f_{\ell}(e,\ud), f_a(e,\ud) [=f_\epsilon(e,\ud)]$ and $f_e(e,\ud)$.
They are plotted as a function of $e$ for $\ud=\pi$ (solid lines),
$\pi/2$ (dashed lines) and $\pi/4$ (dotted lines).
}
\label{fig:factor}
\end{figure}

By numerically integrating Eqs.~(\ref{eq:e_evol}) and (\ref{eq:a_evol}), 
we obtain the evolution of the orbital elements 
according to the planetary mass growth.
In Fig.~\ref{fig:factor}, we plot $f_{\ell}(e,\ud), f_a(e,\ud) [=f_\epsilon(e,\ud)]$ 
and $f_e(e,\ud)$ as a function of $e$ for $\ud=\pi,\pi/2$ and $\pi/4$.
Because $f_e(e,\ud), f_a(e,\ud) > 0$ for any values of
$e$ and $\ud$, $e$ and $a$ monotonically decrease as
the planet grows through accretion of disk gas.
For $\ud = \pi$ (embedded case), $f_\ell$ dominates
the eccentricity damping.
That is, accretion of disk gas with high specific angular momentum
near the apocenter is responsible for the eccentricity damping.
On the other hand, for $\ud = \pi/2$ and $\pi/4$ (deviated cases),
$f_\ell$ is small or negative except for high $e$.
In these cases, the energy dissipation by collision between
incident gas and the planet is responsible for the eccentricity damping
(see section 4.2).

When $e$ becomes small enough, ${\rm d}\log a/{\rm d}\log M$ quickly approaches zero
($f_a\rightarrow 0$).
Thus, the asymptotic values of $a$ are
uniquely determined by the initial values of $e$, $a$, and $\rd$.
In the next section, we show the numerically obtained evolution paths.

\section{Evolution paths of $e$ and $a$}
\subsection{Embedded case}
\label{sec:embedded}

\begin{figure}
\epsscale{1.2}
\plottwo{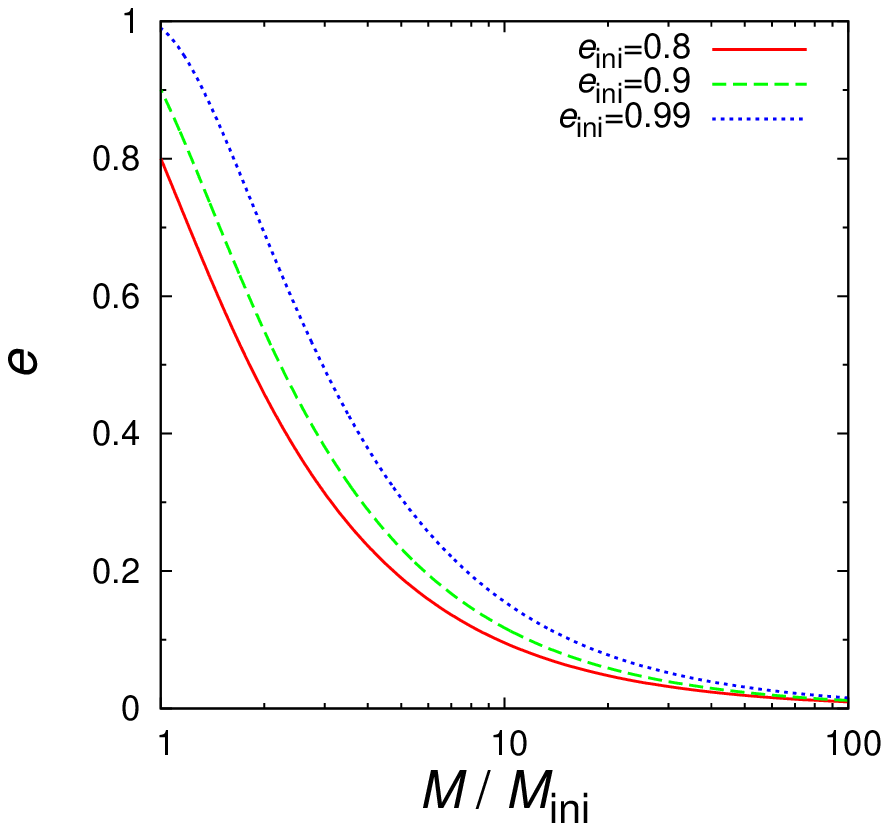}{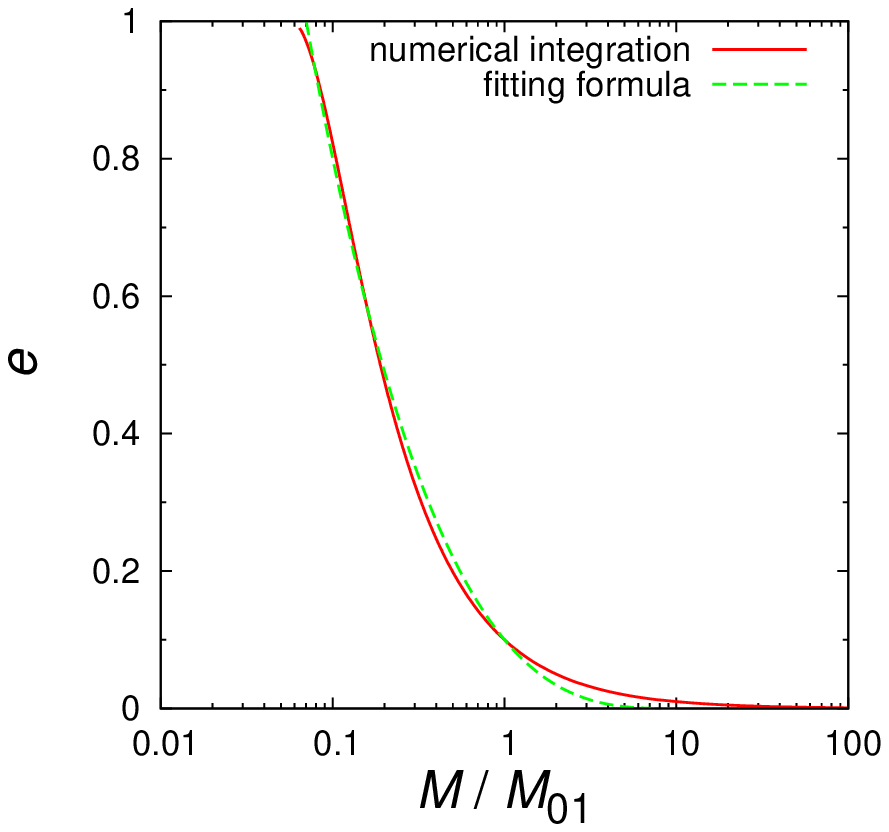}
\caption{
Evolution of $e$ as a function of $M$.
(a) $M$ is scaled by $M_{\rm ini}$ and
$e_{\rm ini}=0.8$ (the solid line), 0.9 (the dashed line) and 0.99 
(the dotted line) are plotted.
(b) $M$ is scaled by $M_{01}$, where $M_{01}$ is
$M$ at $e=0.1$.  In this case, $e$ is uniquely determined by $M/M_{01}$
(the solid line).
The fitting formula, Eq.~(\ref{eq:eM_appr}), which is presented
in section~\ref{sec:pop} is also plotted with the dashed line.
}
\label{fig:e_M}
\end{figure}

\begin{figure}
\epsscale{1.2}
\plottwo{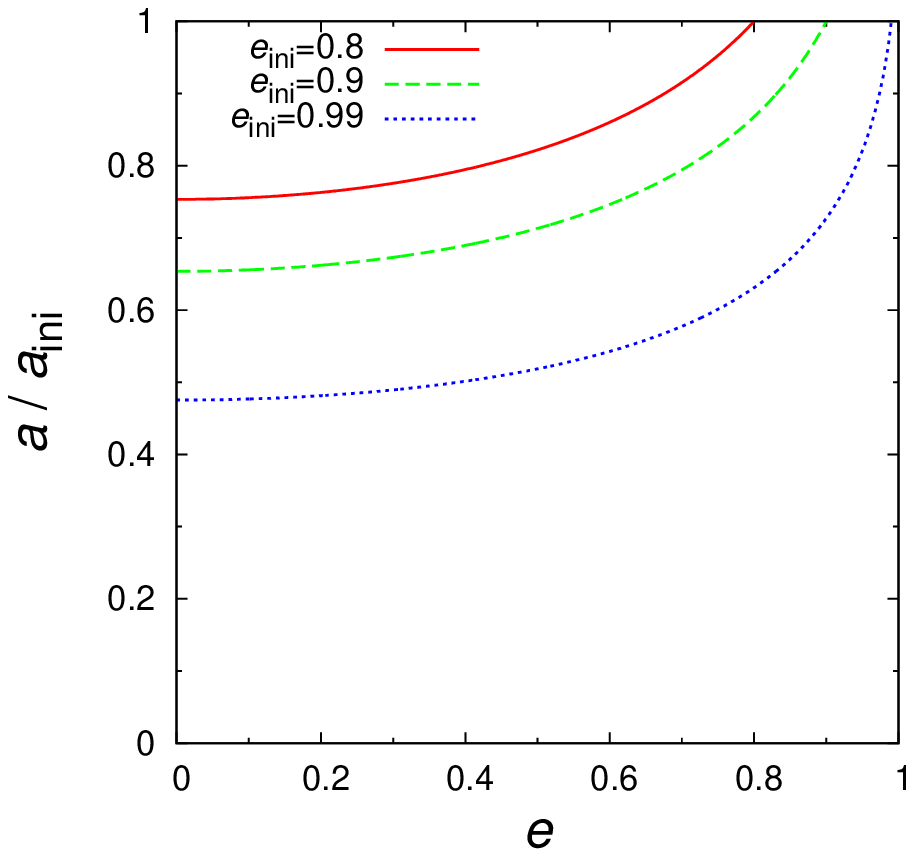}{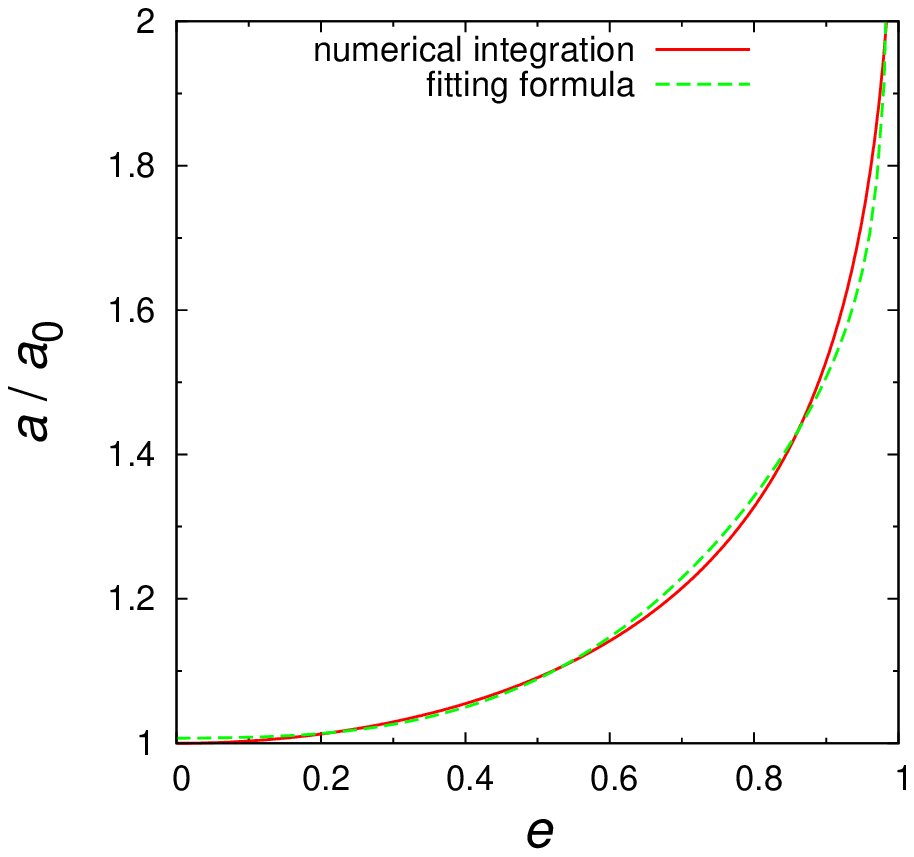}
\caption{
Evolution of $e$ as a function of $a$.
(a) $a$ is scaled by $a_{\rm ini}$ and
$e_{\rm ini}=0.8$ (the solid line), 0.9 (the dashed line) and 0.99 
(the dotted line) are plotted.
(b) $a$ is scaled by $a_{0}$, where $a_{0}$ is
the asymptotic values of $a$ at $e \rightarrow 0$. 
In this case, $e$ is uniquely determined by $a/a_{0}$  (the solid line).
The fitting formula, Eq.~(\ref{eq:ea_appr}), which is presented
in section~\ref{sec:pop} is also plotted with the dashed line.
}
\label{fig:e_a}
\end{figure}

\begin{figure}
\epsscale{1.2}
\plottwo{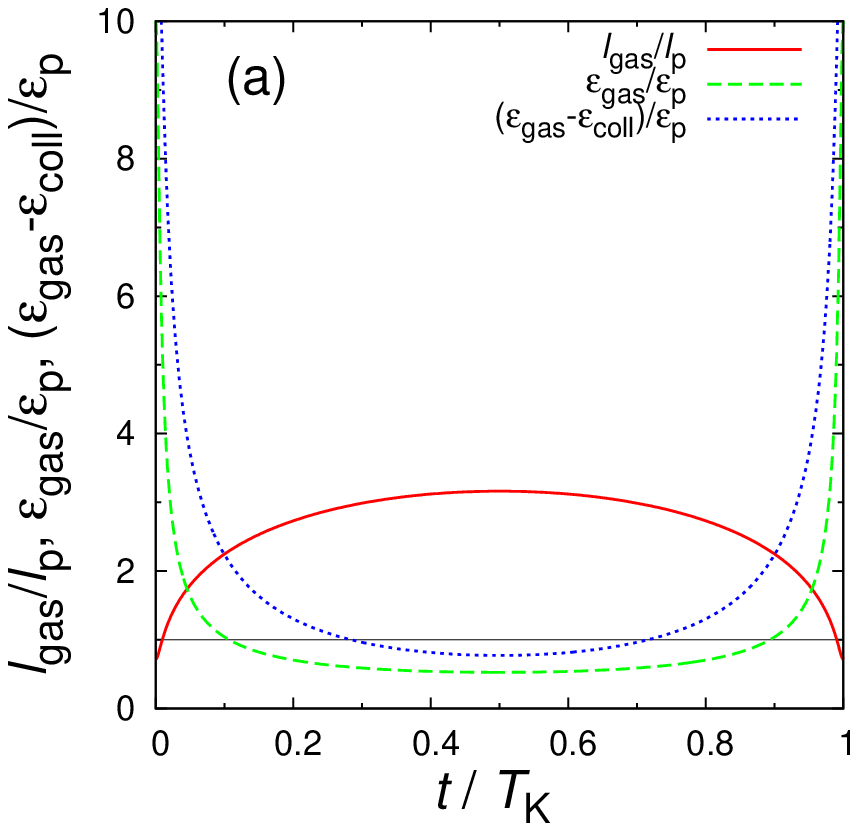}{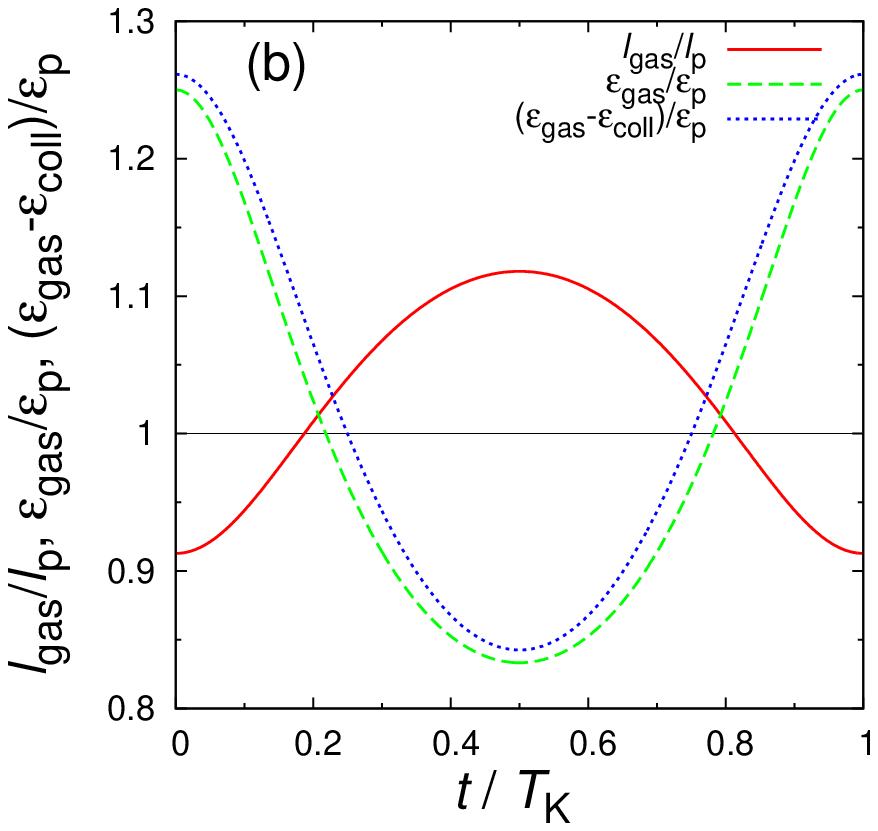}
\caption{
The ratios of specific orbital angular momentum (energy)
of local gas ($\ell_{\rm gas}$, $\epsilon_{\rm gas}$)
to those of the planet ($\ell_{\rm p}$, $\epsilon_{\rm p}$)
as a function of $t/T_{\rm K}$.
For (a) $e = 0.9$ and (b) 0.2,
$\ell_{\rm gas}/\ell_{\rm p}$(the solid lines),
$\epsilon_{\rm gas}/\epsilon_{\rm p}$(the dashed lines)
and $(\epsilon_{\rm gas} - \epsilon_{\rm coll})/\epsilon_{\rm p}$(the dotted lines)
are plotted.
The pericenter and apocenter correspond to $t/T_{\rm K}=0$ and $t/T_{\rm K}=0.5$, respectively.
}
\label{fig:L_E}
\end{figure}

\begin{figure}
\epsscale{.60}
\plotone{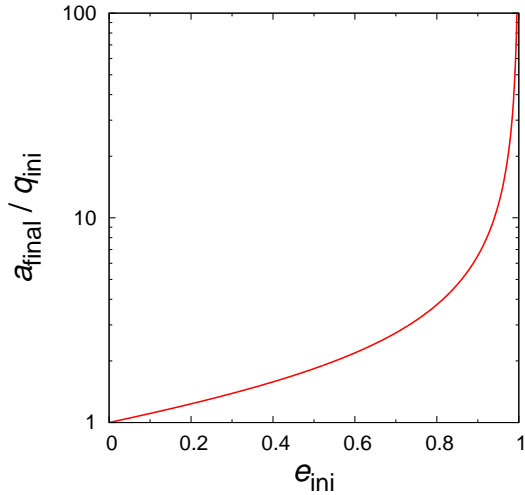}
\caption{
The ratio $a_{\rm final}/q_{\rm ini}$ as a function of $e_{\rm ini}$,
where $a_{\rm final}$ is asymptotic semimajor axis after the orbital circularization
and $q_{\rm ini}$ is initial pericenter distance before the circularization.
}
\label{fig:q_e}
\end{figure}

\begin{figure}
\epsscale{.60}
\plotone{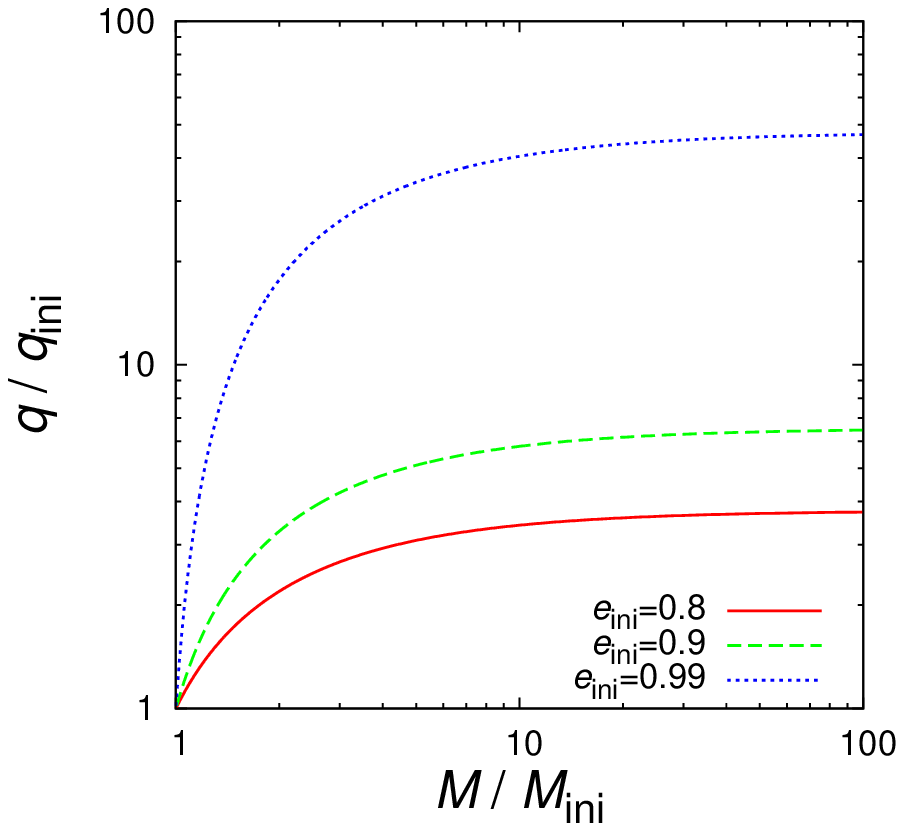}
\caption{
The evolution of $q/q_{\rm ini}$ as a function of $M/M_{\rm ini}$.
$e_{\rm ini}=0.8$ (the solid line), 0.9 (the dashed line) and 0.99 
(the dotted line) are plotted.
}
\label{fig:q_M}
\end{figure}

First, we consider embedded case, that is,
whole parts of a planetary orbit is embedded in the disk.
Since equation~(\ref{eq:A_e}) is independent of multiplication
of $M$ by a constant factor,
we can adopt a scaled quantity $M/M_{\rm ini}$
as a variable,
where $M_{\rm ini}$ is the initial value of $M$.
Then, Eq.~(\ref{eq:A_e}) includes only $e$ and $M/M_{\rm ini}$,
so that the evolution of $e$ is uniquely 
given as a function of $M/M_{\rm ini}$ 
for any initial values of eccentricity ($e_{\rm ini}$).
The evolutional paths for representative values of $e_{\rm ini}$
that obtained by numerical integration of Eq.~(\ref{eq:A_e}) are
shown in Figure~\ref{fig:e_M}a.
This figure shows that $e$ decreases to values
below $0.2e_{\rm ini}$ when $M$ attains $10M_{\rm ini}$.
Since a typical core mass to start runaway gas accretion
is $\sim 10M_\oplus$, it means that $e$ is reduced to 
values smaller than 0.2
when the planet acquires Saturnian mass ($\sim 100M_\oplus$),
even if its initial orbit was close to a parabolic orbit ($e_{\rm ini} \sim 1$).
If the mass is scaled by that at $e=0.1$, denoted by $M_{01}$,
$e$ is uniquely determined by $M/M_{01}$.
Figure~\ref{fig:e_M}b shows the self-similar solution 
of $e$ as a function of $M/M_{01}$ (the solid curve). 
The fitting formula given by Eq.~(\ref{eq:eM_appr}), which is presented
in section~\ref{sec:pop} is also plotted with the dashed curve.

Since Eq.~(\ref{eq:A_ea}) has a similar structure to Eq.~(\ref{eq:A_e}),
the evolution of $e$ is uniquely 
given as a function of $a/a_{\rm ini}$. 
Figure~\ref{fig:e_a}a shows
the evolutional paths on the $a$-$e$ plane
for representative values of $e_{\rm ini}$.
Because both $e$ and $a$ keep decreasing,
the evolution starts at the right end 
and moves leftward.
Damping of $e$ is dominated over that of $a$ except 
for $e \simeq 1$.
Even if $e_{\rm ini} = 0.99$, the asymptotic 
value of $a$ for $e \rightarrow 0$, which we denote as $a_{\rm final}$,
is as much as $\sim 0.48a_{\rm ini}$.
As in the case of $e$-$M$ relation,
if the semimajor axis is scaled by that at a specific value of $e$,
the evolution of $e$ is expressed by a single curve,
irrespective of $e_{\rm ini}$ and $a_{\rm ini}$.
Figure~\ref{fig:e_a}b shows the self-similar relation, 
$a/a_{0}$ as a function of $e$ (the solid curve),
where $a_{0}$ is $a$ at $e=0$.
The fitting formula given by Eq.~(\ref{eq:ea_appr}), which is presented
in section~\ref{sec:pop} is also plotted with the dashed curve.
This plot clearly shows that damping of $a$ is much smaller than
that of $e$: for the damping of $e$ from $\sim 1$ to 0, 
$a$ is decreased by $\sim 50\%$, and for that from 0.8 to 0,
the decrease in $a$ is only $\sim 30\%$.

Figure~\ref{fig:L_E}a shows 
specific orbital angular momentum and energy
of local gas ($\ell_{\rm gas}$ and $\epsilon_{\rm gas}$)
scaled by those of the planet ($\ell_{\rm p}$ and $\epsilon_{\rm p}$)
for $e=0.9$ as functions of $t/T_{\rm K}$.
The pericenter passage is at $t/T_{\rm K}=0$ and $t/T_{\rm K}=1$,
and the apocenter passage is at $t/T_{\rm K}=0.5$, respectively.
Because the orbit is highly eccentric,
in most of time of an orbital period, 
$\ell_{\rm gas} > \ell_{\rm p}$ and $\epsilon_{\rm gas} > \epsilon_{\rm p}$
(since the energy is negative, $| \epsilon_{\rm gas} | < | \epsilon_{\rm p}|$)
except in the regions close to pericenter ($t/T_{\rm K}=0$ and $t/T_{\rm K}=1$).
As a result, an orbit-averaged value of $\ell_{\rm gas}/\ell_{\rm p}$ is
considerably larger than unity (in this case, it is 
$\langle \ell_{\rm gas}/\ell_{\rm p} \rangle =f_\ell(0.9,\pi)+1=2.66$)
and the specific angular momentum of the planet increases
through accretion of disk gas.

On the other hand, 
it is analytically shown that an orbit-averaged value of 
$\epsilon_{\rm gas}/\epsilon_{\rm p}$ is unity,
since 
\begin{equation}
\biggl\langle \frac{\epsilon_{\rm gas}}{\epsilon_{\rm p}} \biggr\rangle 
=\frac{1}{T_{\rm K}} \int_{-T_{\rm K}/2}^{T_{\rm K}/2} 
\bra \frac{\epsilon_{\rm gas}}{\epsilon_{\rm p}}\ket {\rm d} t
=\frac{1}{T_{\rm K}} \int_{-T_{\rm K}/2}^{T_{\rm K}/2} 
\bra \frac{r}{a} \ket^{-1} {\rm d} t = 
f_{-1}(e,\pi) =
\frac{\pi}{\pi - e \sin \pi} = 1.
\end{equation}
Although $| \epsilon_{\rm gas}|  < | \epsilon_{\rm p}| $ in most of time,
$|\epsilon_{\rm gas}|$ is much larger than $ | \epsilon_{\rm p}| $
near the pericenter passage (Fig.~\ref{fig:L_E}a).
The contribution of large $|\epsilon_{\rm gas}|$ compensates for 
the excess energy accretion in outer regions.
But, the orbital energy of the planet decreases 
because it is also contributed by the collisional energy dissipation 
$\epsilon_{\rm coll}(=\vrel^2/2)$.
Equation~(\ref{eq:Delta_e}) shows that the collisional energy
dissipation also damps $e$.
If the collisional energy dissipation is neglected, $a$ is conserved and $e$ is damped
by the accretion of higher specific angular momentum gas.
With the effect of the collisional energy dissipation, 
damping of $e$ is faster and $a$ is also damped
while the $a$-damping is slower than the $e$-damping.

From these relations, together with $\epsilon_{\rm p} = -GM_\ast/2a$ and $\ell_{\rm p} = \sqrt{GM_\ast a(1-e^2)}$, 
it is readily found that both $a$ and $e$ always decrease 
in the case of constant gas accretion rate.
Figure~\ref{fig:L_E}b shows
$\ell_{\rm gas}/\ell_{\rm p}$ 
and $\epsilon_{\rm gas}/\epsilon_{\rm p}$ at $e=0.2$.
In this case, the integrals are more symmetric about 
$\ell_{\rm gas}/\ell_{\rm p}=1$ 
and $\epsilon_{\rm gas}/\epsilon_{\rm p}=1$. 
Thereby, when $e$ is reduced to $\la 0.2$,
the orbit-averaged values of $\ell_{\rm gas}/\ell_{\rm p}$ 
and $\epsilon_{\rm gas}/\epsilon_{\rm p}$ are nearly unity
and the decrease in $e$ and $a$ due to planetary mass growth slows down.

Initial pericenter distance $q_{\rm ini}$
of the core's orbit before the $e$-damping process would correspond to
the original semimajor axis of the core ($a_{\rm ori}$) before the core
was scattered by a gas giant, which may be $\sim 1$--10AU.
Figure~\ref{fig:q_e} shows asymptotic semimajor axis $a_{\rm final}$
scaled by $q_{\rm ini}$.
Because $a_{\rm final}$ is the final semimajor axis of a gas giant
formed from a scattered core after $e$ is damped, 
$a_{\rm final}/q_{\rm ini}$ indicates an efficiency to send a planet
to outer regions.
In this figure, we find that 
a core originally at inner region ($a_{\rm ori} \sim q_{\rm ini} \sim$ 10AU)
become a gas giant with large radius ($a_{\rm final} \ga 30$AU when 
$e_{\rm ini} \ga 0.73$, and $a_{\rm final} \ga 100$AU when $e_{\rm ini} \ga 0.94$). 

Figure~\ref{fig:q_M} shows the evolution of pericenter distance 
scaled by initial one, $q/q_{\rm ini}$, due to planetary mass growth,
for representative values of $e_{\rm ini}$.
It is shown that $q/q_{\rm ini}$ quickly increases, which justfies
our assumption that the scattered planet becomes quickly isolated and the further perturbations
from the gas giant in inner region are neglected.

\begin{figure}
\epsscale{.60}
\plotone{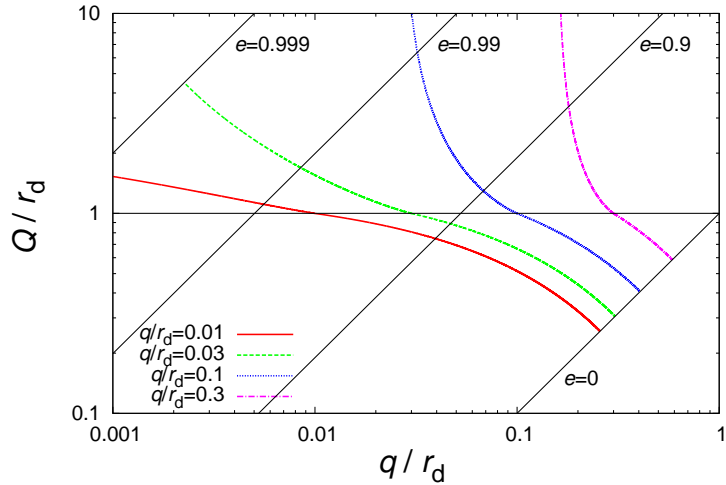}
\caption{
The evolution of apocenter distance $Q$ and pericenter distance $q$.
Both are scaled by the disk size $\rd$.
The evolution paths are parameterized by the values of $q/\rd$
at $Q = \rd$ at which the deviated-case evolution is switched to embedded-case evolution;
the paths with $q/\rd |_{Q = \rd}$ =  0.01 (the solid line), 0.03 (the dashed line), 0.1 (the dotted line) and 0.3 (the dot-dashed line) 
are plotted.
}
\label{fig:Qq_B}
\end{figure}
\begin{figure}
\epsscale{.60}
\plotone{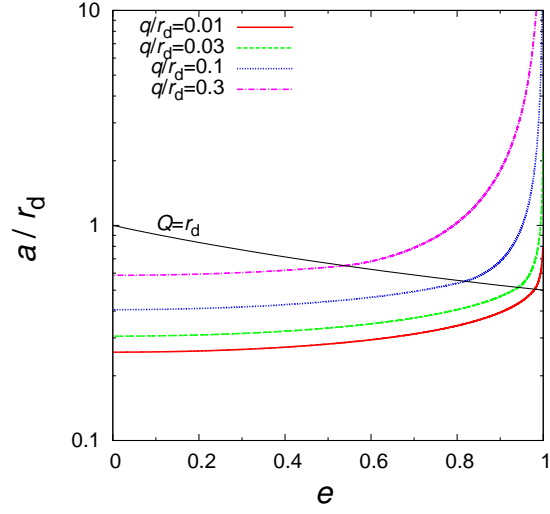}
\caption{
The evolutions of $e$ and $a/\rd$ 
corresponding to the solutions in Fig.~\ref{fig:Qq_B}.
}
\label{fig:ard_e}
\end{figure}

\begin{figure}
\epsscale{.60}
\plotone{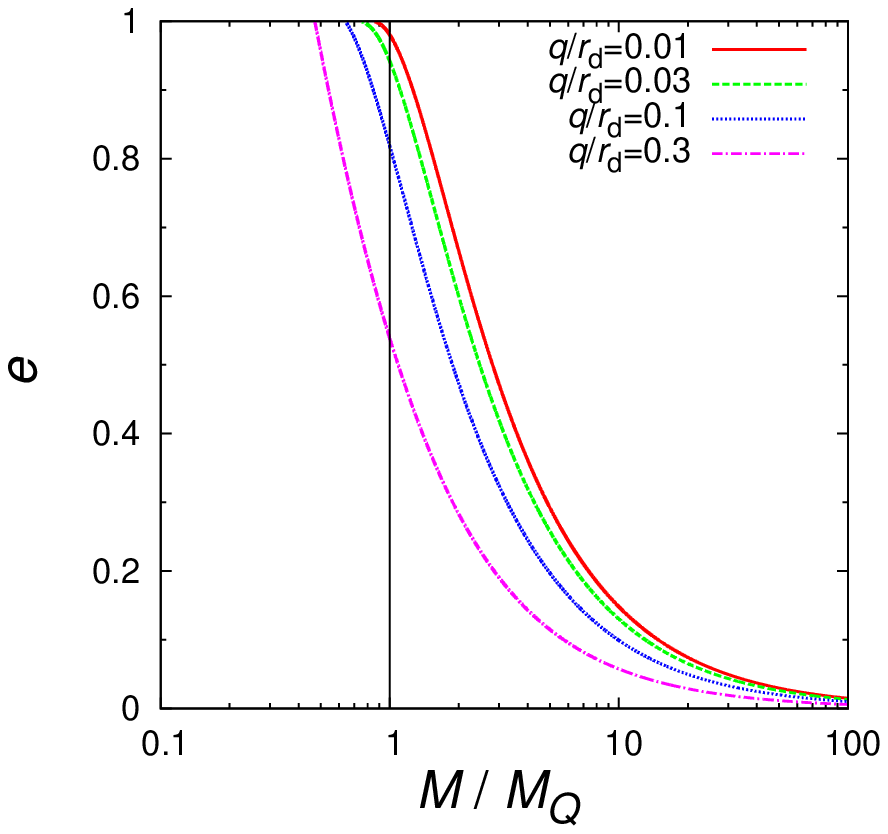}
\caption{
The evolution of $e$
as a function of $M/M_{Q}$
corresponding to the solutions in Fig.~\ref{fig:Qq_B},
where $M_{Q}$ is the planetary mass at $Q=\rd$.
}
\label{fig:e_MMQ}
\end{figure}

\subsection{Deviated case}

Next, we consider deviated case in which $Q > \rd$.
Since in this case, gas accretion is halted at $r > \rd$,
the planet cannot accrete gas with higher specific angular momentum and energy,
resulting in smaller $f_\ell$ and larger $f_\epsilon(=f_a)$ (see Fig.~\ref{fig:factor}).
The increase of $f_\epsilon(=f_a)$ is more effective than the decrease of $f_\ell$, 
so $f_e$ is larger.
Thus, both $e$ and $a$ dampings in the deviated case are more efficient than in the embedded case.

Since there is a characteristic length $\rd$, 
a self-similar solution like Fig.~\ref{fig:e_a}b does not exist.
However, it is clear that evolutions of $Q/\rd$ and $q/\rd$ should be the same 
for the same initial values.
Figure~\ref{fig:Qq_B} shows the evolutions of $Q/\rd$ and $q/\rd$.
The evolutions are to the right-down direction.
The evolutions in deviated case correspond to those in the region of $Q > \rd$.
We also added following embedded evolutions in the region of $Q < \rd$.
Because the evolution paths do not cross each other,
we can parameterize the evolution paths with one parameter.
In Fig.~\ref{fig:Qq_B}, we used the value of
$q/\rd$ at the time when $Q$ is reduced to be $\rd$,
as the parameter.

The evolutions of $e$ and $a$ corresponding to the
solutions in Figure~\ref{fig:Qq_B} are plotted in Figure~\ref{fig:ard_e}.
The orbital evolution is toward the left-down direction.
This figure shows that in the early phase of $Q > \rd$, 
the semimajor axis is predominantly damped.
Note that even in this
phase where $a$ is rapidly reduced,  Figure~\ref{fig:Qq_B} shows that
$q$ is increased so quickly that
the planet becomes isolated from the perturbing gas giant.
For $e_{\rm ini} \sim 1$,
$a$ is damped by order of magnitude until $Q \sim a(1+e) \sim 2a$
is reduced to $\sim \rd$.
In the following embedded phase of $Q < \rd$, however, 
$a$ is reduced at most by a factor of 2,
as we showed.
Thus, in this case, $a_{\rm final} \sim \rd/4$, independent of the
values of $a_{\rm ini}$, as long as $e_{\rm ini} \sim 1$.
In other words, we can infer the values of $\rd$ from $a_{\rm final}$.

Figure~\ref{fig:e_MMQ} shows the evolution of $e$
as a function of the planetary mass $M/M_{Q}$,
where $M_{Q}$ is the planetary mass at $Q=\rd$.
In the deviated phase, the $e$-damping is more efficient than in the embedded phase, although it is slightly slower because of high eccentricity.
%

\subsection{Effect of shock}

We have considered the energy dissipation by collision
between incident gas flow and the planet.
The dissipation is needed to bind the gas around the planet
and it accelerates eccentricity damping as we showed.
The dissipation should occur through bow shock in front of the planet.
The shock not only causes the energy dissipation but also 
makes the relative velocity lower.
So far, we have neglected the relative velocity damping by shock,
which should weaken the eccentricity and semimajor axis damping.
Here we evaluate the effect of the shock using a simple 1D model.
Full 2D or 3D hydrodynamical simulations will be done in a separate paper. 

The simple 1D model we use is as follows.
The ratio of post-shock velocity ($s v_{\rm rel}$) to pre-shock one ($v_{\rm rel}$) is
\begin{equation}
s =\frac{(\gamma-1){\cal M}^2+2}{(\gamma+1){\cal M}^2}
\simeq \frac{1}{4} \left( 1+ \frac{3}{{\cal M}^2} \right),
\end{equation}
where 
$\gamma$ is the specific heat ratio ($\gamma=5/3$ in the monatomic molecule) 
and ${\cal M}$ is Mach number 
for pre-shock gas flow, which is given by
\begin{equation}
{\cal M}=\frac{\vrel}{c_s}
= 30 \left( \frac{r}{1{\rm AU}} \right)^{-1/4} \left( 3 - \frac{r}{a} - 2\sqrt{\frac{a}{r} (1-e^2) } \right)^{1/2},
\end{equation}
where we used an optically thin disk temperature, $T = 280(r/1{\rm AU})^{-1/2}{\rm K}$, for evaluation of sound velocity $c_s$.
For subsonic case (${\cal M} <1$), $s = 1$.
The radial and tangential components of velocity of post-shock gas ($u_r, u_\phi$) 
and those of pre-shock gas ($0, v_{\rm K}$) are related to those of planet ($v_r, v_\phi$), which
are given by Eqs.(\ref{eq:v_r}) and (\ref{eq:v_phi}), as
\begin{align}
u_r - v_r &= s (0 - v_r), \\
u_\phi - v_\phi &= s (v_{\rm K} - v_\phi).
\end{align}
Then, the integrants of Eqs.~(\ref{eq:f_ell}) and (\ref{eq:f_epsilon}) 
for $f_\ell$ and $f_\epsilon$ are
\begin{align}
\frac{l_{\rm gas}}{l_{\rm p}} -1
&= \frac{r u_\phi}{r v_\phi} -1
= s \left( \frac{r v_{\rm K}}{r v_\phi} -1 \right)
= s \left( \frac{l_{\rm gas,0}}{l_{\rm p}} -1 \right), \label{eq:delta_l}\\
\frac{\epsilon_{\rm gas}}{\epsilon_{\rm p}}  - \frac{\epsilon_{\rm coll}}{\epsilon_{\rm p}} -1
&= \frac{(u_r^2+u_\phi^2)/2 - GM_\ast/r - s^2 \vrel^2/2} 
{(v_r^2+v_\phi^2)/2 - GM_\ast/r} -1 \nonumber\\
&= s \left( \frac{v_{\rm K}^2/2 - GM_\ast/r - \vrel^2/2}{(v_r^2+v_\phi^2)/2 - GM_\ast/r}  -1 \right) = s \left( \frac{\epsilon_{\rm gas,0}}{\epsilon_{\rm p}}  - \frac{\epsilon_{\rm coll,0}}{\epsilon_{\rm p}} -1 \right),
\label{eq:delta_eps}
\end{align}
where quantities with subscript ',0' means those for unperturbed gas flow
neglecting shock
($l_{\rm gas,0}=\sqrt{GM_\ast r}$, $\epsilon_{\rm gas,0}=-GM_\ast/2r$, 
$\epsilon_{\rm coll,0}=\vrel^2/2$).

\begin{figure}
\epsscale{.60}
\plotone{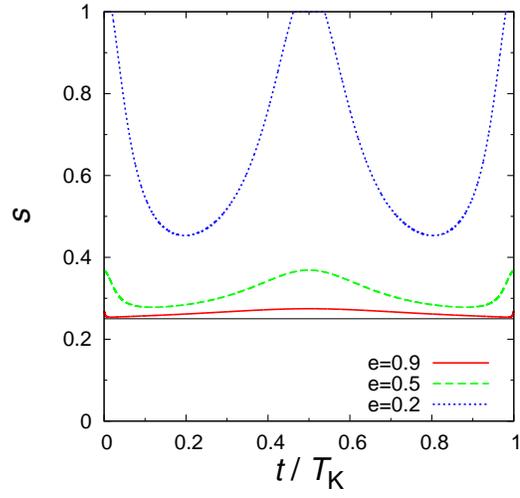}
\caption{
The time dependence of $s$ in one orbit where $s$ is the ratio of velocity of after and before shock.
$e=0.9$ (the solid line), 0.5 (the dashed line) and 0.2 (the dotted line) are plotted.
The pericenter and apocenter correspond to $t/T_{\rm K}=0$ and $t/T_{\rm K}=0.5$, respectively.
We assume an optically thin disk temperature, $T = 280(r/1{\rm AU})^{-1/2}{\rm K}$, and $a = 100$AU.
}
\label{fig:shock_s}
\end{figure}

As we showed in the previous subsections, the integrations for $f_\ell$ and $f_\epsilon$ 
(Eqs.~(\ref{eq:f_ell}) and (\ref{eq:f_epsilon})) can be analytically done in the case 
neglecting the damping of the relative velocity (equivalently, $s=1$).
However, since in the present case, $s$ varies along the orbit (Fig.~\ref{fig:shock_s}),
we integrate Eqs.~(\ref{eq:f_ell}) and (\ref{eq:f_epsilon}) numerically.
Here we consider embedded case.
Since $f_\ell$ and $f_\epsilon$ depend on $a$ through $s$ in the
present case, we assume $a =100$AU ($a$ is variable when we consider the evolution paths).
Note that evolution paths in the $a$-$e$ plane is the same as those in the case
without the relative velocity damping.
In Eq.~(\ref{eq:A_ea}),  ${\rm d}\log a/{\rm d}e$, is given 
approximately by orbit-averaged $f_a$ and $f_e$.
But, more exactly, ${\rm d}\log a/{\rm d}e$ must be integrated every time.
From Eqs.~(\ref{eq:delta_l}) and (\ref{eq:delta_eps}), it is apparent that $s$ 
completely cancels and ${\rm d}\log a/{\rm d}e$ is the same.
 
\begin{figure}
\includegraphics[scale=0.7]{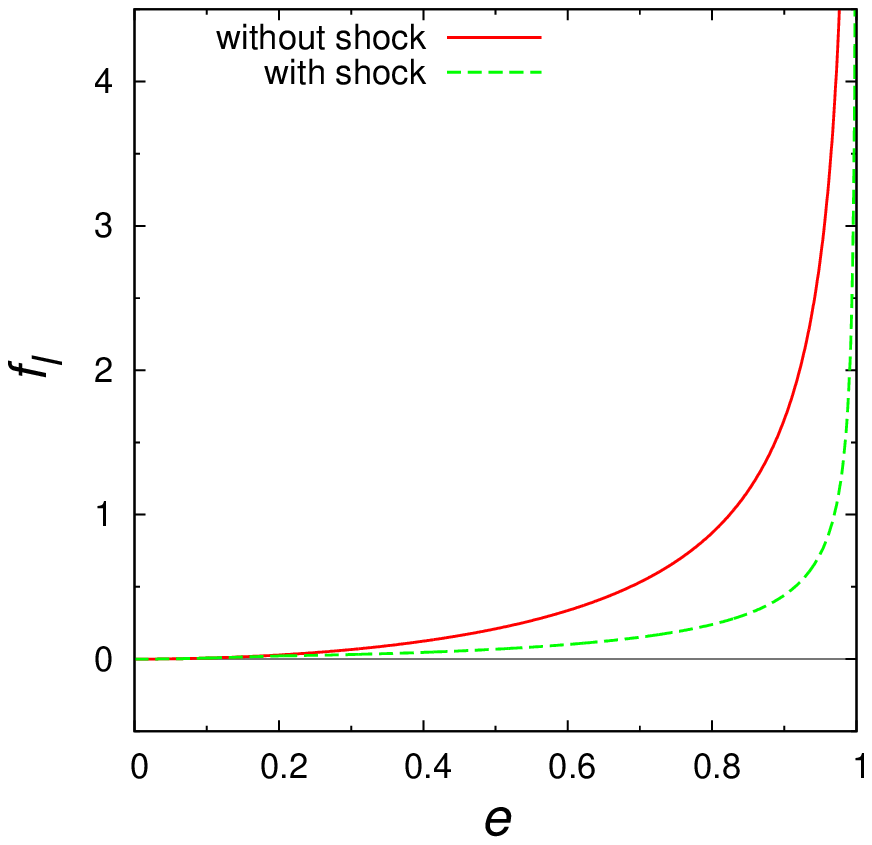}
\includegraphics[scale=0.7]{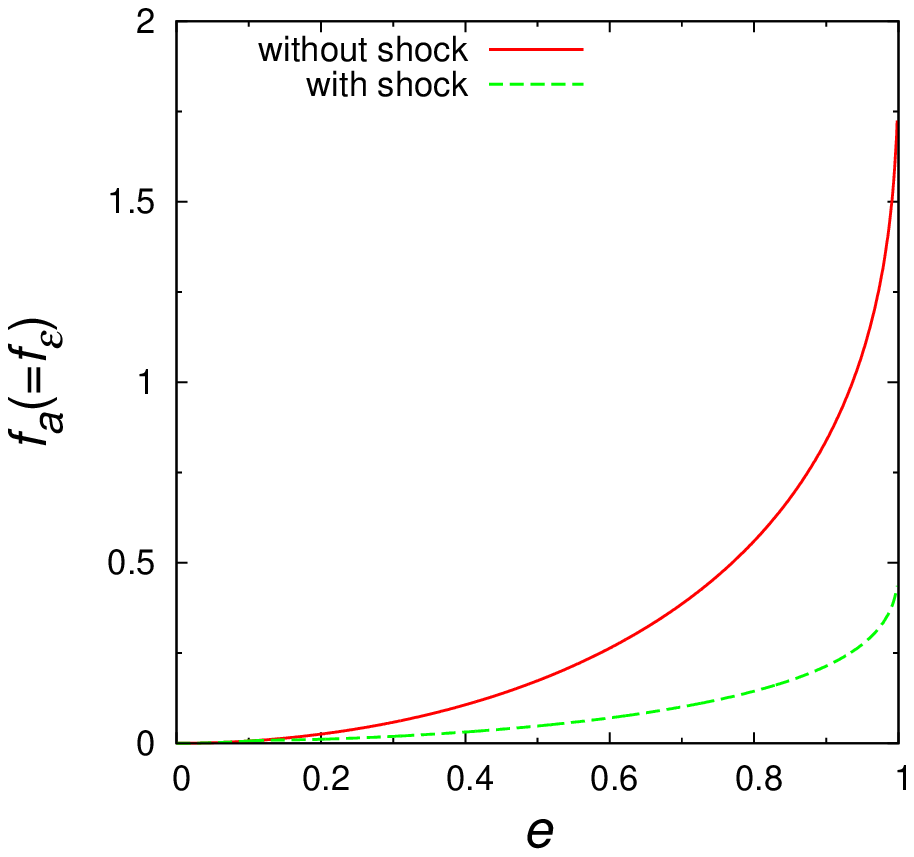}
\includegraphics[scale=0.7]{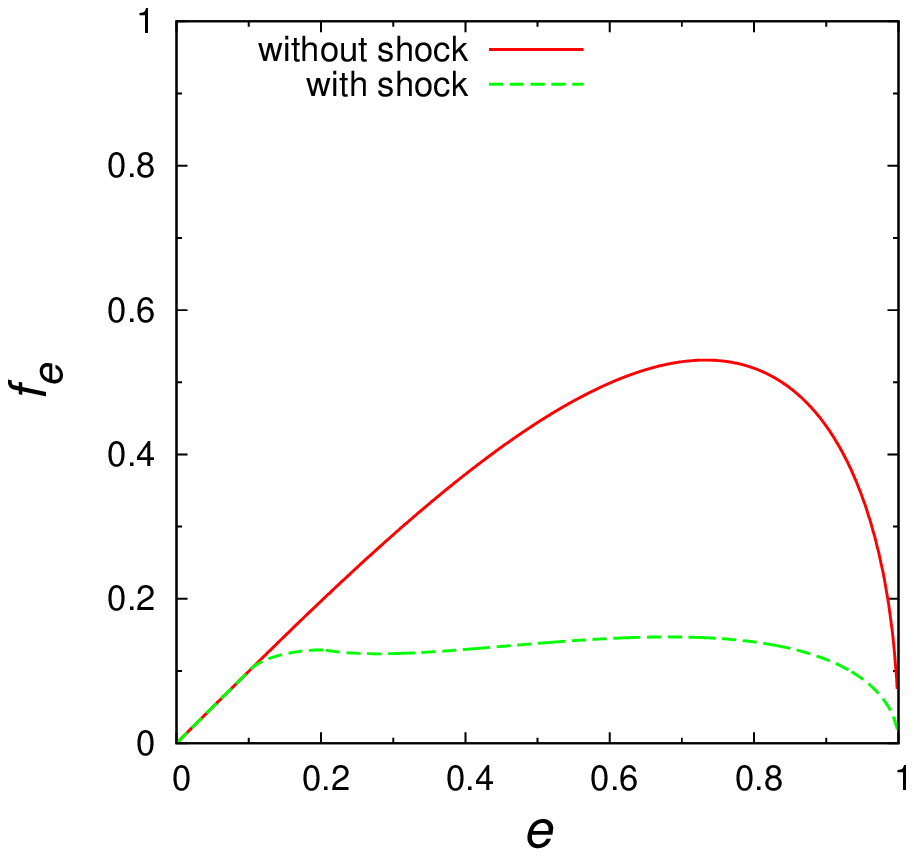}
\caption{
The functions $f_\ell, f_a (=f_\epsilon)$ and $f_e$. 
The solid lines and dotted lines represent the functions without shock 
and with shock, respectively.
We assume $a = 100$AU in shock case.
}
\label{fig:f_shock}
\end{figure}

\begin{figure}
\includegraphics[scale=0.7]{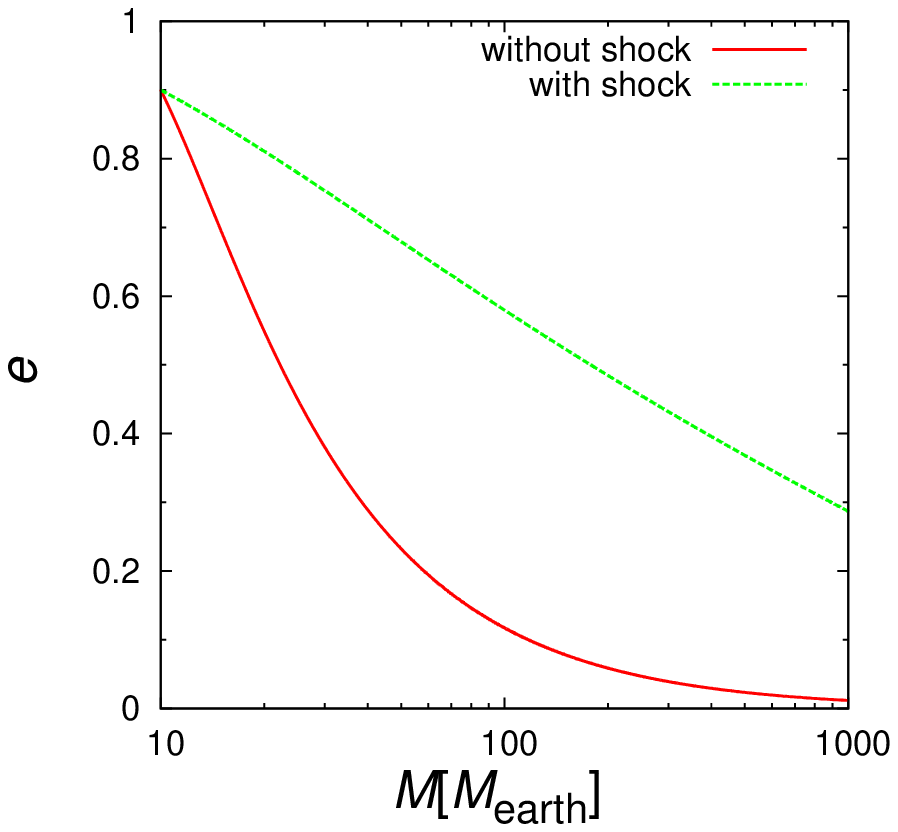}
\includegraphics[scale=0.7]{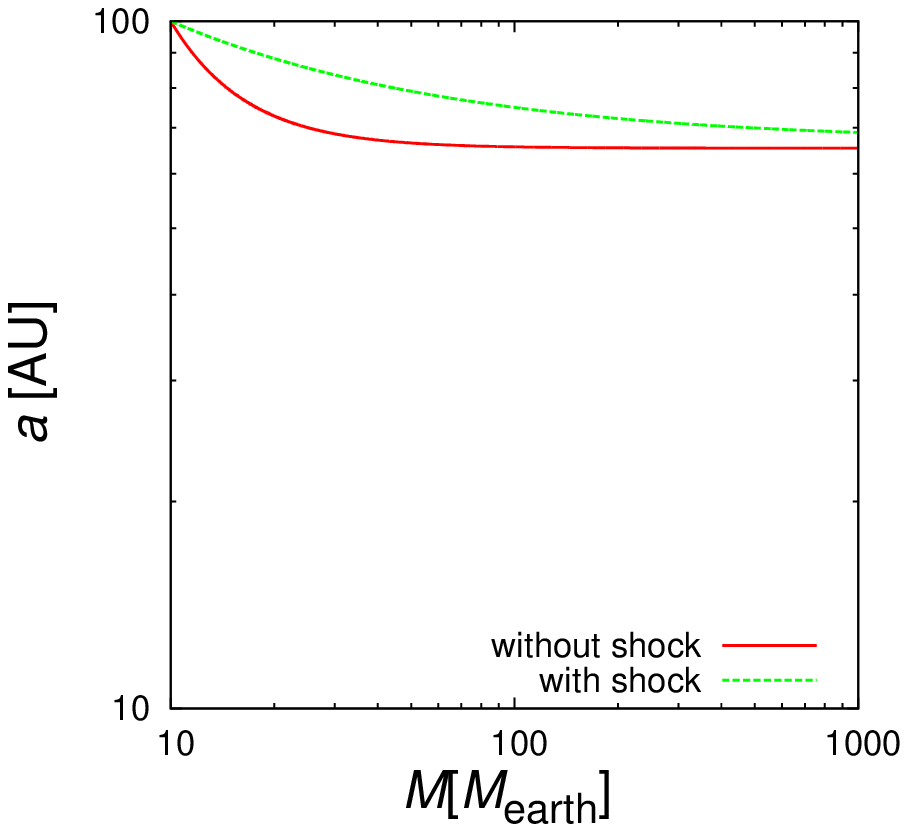}
\includegraphics[scale=0.7]{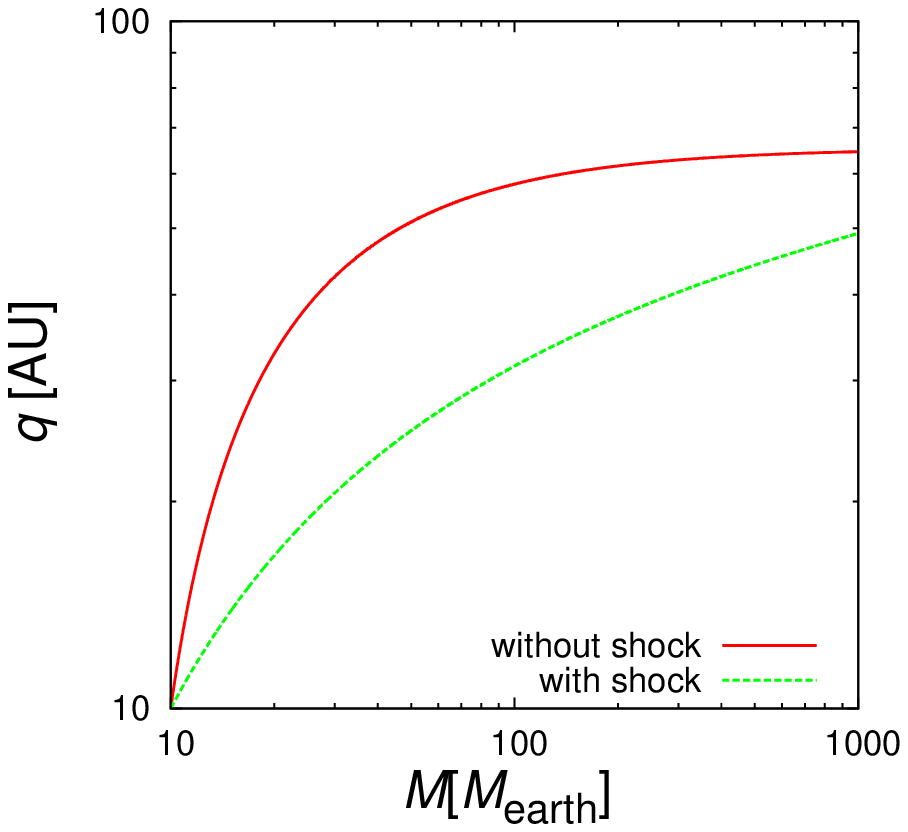}
\caption{
The evolution paths on the $e$-$M$, $a$-$M$, and $q$-$M$ planes.
The initial conditions are $q_{ini}=10$AU, $e_{ini}=0.9$ and $M_{ini}=10M_\oplus$, respectively.
The solid lines and dotted lines represent the functions without shock 
and with shock, respectively.
}
\label{fig:shock_evol}
\end{figure}

The functions $f_\ell, f_a(=f_\epsilon)$ and $f_e$ in the case with shock 
are compared with those without shock in Fig.~\ref{fig:f_shock}.
The effect of shock lowers all the functions.
Accordingly, both $e$ and $a$ dampings are slowed down, while
the evolution paths on the $a$-$e$ plane do not change.
The evolution paths on the $e$-$M$, $a$-$M$, and $q$-$M$ planes are 
shown in Fig.~\ref{fig:shock_evol}.
The initial conditions are $q_{\rm ini}=10$AU, $e_{\rm ini}=0.9$ and $M_{\rm ini}=10M_\oplus$, respectively.
While $e$ declines to values $< 0.2$ at $M_{\rm p}\sim 50 M_\oplus$ in non-shock case,
it does not becomes $< 0.2$ until $M_{\rm p} \sim 3000 M_\oplus$ in shock case.
However, since the masses of direct-imaged planets are relatively large ($\sim 10 M_{\rm J}$),
the orbital circularization is still effective. 
The $a$ damping is also slowed down, but the semimajor axis at $M_{\rm p} \sim 10 M_{\rm J}$
is not significantly larger than that in non-shock case.
The pericenter distance $q$ is still quickly increased, so that the assumption that the scattered planet becomes quickly isolated and the further perturbations from the gas giant in inner region are neglected is justified.
Thus, although  
the eccentricity damping is less efficient, 
formation of distant gas giants in nearly circular orbits is not significantly inhibited by the effect of shock.

\clearpage
\section{Fitting formulas and population synthesis simulation}
\label{sec:pop} 

The self-similar solution in Fig.~\ref{fig:e_M}b 
can be approximately fitted by
\begin{equation}
e  \simeq 0.1\left[1- \log_{10}\left(\frac{M}{M_{01}}\right)\right]^3.
\label{eq:eM_appr} 
\end{equation}
where $M_{01}$ is $M$ at $e=0.1$.
Then, for any given $e_{\rm ini}$ and $M_{\rm ini}$, $e$ for $M (> M_{\rm ini})$
are evaluated by
\begin{align}
e & \simeq 0.1\left[1 - \log_{10}\left(\frac{M_{\rm ini}}{M_{01}}\right) - \log_{10}\left(\frac{M}{M_{\rm ini}}\right)\right]^3 \nonumber\\
   & = 0.1\left[\left(\frac{e_{\rm ini}}{0.1}\right)^{1/3} - \log_{10}\left(\frac{M}{M_{\rm ini}}\right) \right]^3.
\label{eq:eM_appr2}
\end{align}
Because this is an approximate formula,
Eq.~(\ref{eq:eM_appr2}) can be negative for large values of $M/M_{\rm ini}$.
In such cases, we set $e=0$, because $e$ has very small values 
in the exact solution.
On the other hand, 
the self-similar solution in Fig.~\ref{fig:e_a}b
can be fitted by
\begin{equation}
 \frac{a}{a_0} \simeq 1 + 0.6 e^3 + \frac{0.007}{1-e}.
\label{eq:ea_appr}
\end{equation}
where $a_{0}$ is the asymptotic values of $a$ at $e \rightarrow 0$.
Since a similar relation holds for $a_{\rm ini}$ and $e_{\rm ini}$,
\begin{equation}
 \frac{a}{a_{\rm ini}} \simeq \frac{1 + 0.6 e^3 + 0.007/(1-e)}{1 + 0.6 e_{\rm ini}^3 + 0.007/(1-e_{\rm ini})}.
\label{eq:ea_appr2}
\end{equation}

\begin{figure}[btp]
\epsscale{1.0}       
\plotone{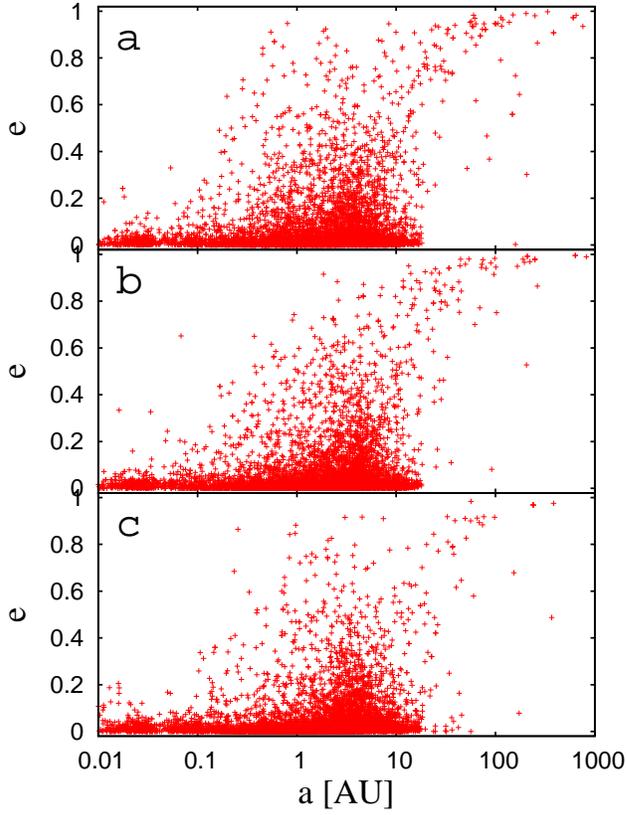} 
\caption{
The distributions of $e$ and $a$ of gas giant planets
around solar-type stars, obtained by 
population synthesis calculations with 
similar parameters to those of the results in Fig.~7 of \citet{ida13}.
For the details of calculations, see \citet{ida13}.
In panel a, neither the dynamical friction to cores nor the damping 
via gas accretion is included.
In panel b, only the dynamical friction is included.
Both effects are included in panel c.
  }
\label{fig:ea3}
\end{figure}
  
In the population synthesis simulation,
when a core with mass $M_c$ closely encounters with a gas giant,
eccentricity and semimajor axis that are excited by the scattering are evaluated
with a Monte-Carlo procedure
 \citep[see, e.g.,][]{ida13}.
We set these eccentricity, semimajor axis, and $M_c$ as
$e_{\rm ini}$, $a_{\rm ini}$, and $M_{\rm ini}$ in the above
equations, respectively.
The mass growth of the planet due to gas accretion after the scattering
is also calculated in the population synthesis simulation.
The mass growth is truncated when disk gas is severely depleted or a clear gap 
along the planetary orbit is opened \citep[see, e.g.,][]{ida04a,ida13}.
From Eq.~(\ref{eq:eM_appr2}), the value of $e$ when the planet mass increases to $M$ 
can be derived from $e_{\rm ini}$ and $M_{\rm ini}$.
The semimajor axis $a$ at $M$ is derived from $e_{\rm ini}$, $e$ and $a_{\rm ini}$
from Eq.~(\ref{eq:ea_appr2}).
Note that
\citet{ida13} simply assumed $e=0$ and $a=a_{\rm ini}$.

Figures~\ref{fig:ea3} show the $e$-$a$ distributions of gas giant planets
around solar-type stars, obtained by a population synthesis
calculation with similar parameters to those of the results in Fig.~7 of \citet{ida13}.
Note that rocky and icy planets with smaller masses are not plotted here.
In panel a, neither the dynamical friction to cores nor the damping 
via gas accretion is included.
Most of giant planets at $\ga 30$AU have large eccentricities, because
they suffered strong gravitational scattering by other giants.
In panel b, only the dynamical friction is included.
Eccentricities of small number of planets are damped, but the effects are not significant.  
On the other hand, both effects are included in panel c.
Eccentricities are damped to values below 0.2 for $\sim 30\%$ of giant planets.
However, since the damping is not as efficient as the simple treatment in \citet{ida13},
the fraction of systems that have gas giants with $a>30$AU and $e<0.2$
is $\sim 0.1\%$, which is smaller than the probability
($\sim 0.4\%$) in Fig.~7 of \citet{ida13}.
It is a future problem to check if such low fraction is consistent with
direct imaging surveys.
As already pointed out in \citet{ida13}, the formation rate
of high eccentricity gas giants at $\sim O(1)$AU is lower in the theoretical prediction
than that found by radial velocity surveys.
If the theoretical prediction is improved so that more frequent formation of
high eccentricity gas giants
is reproduced, the theoretically predicted fraction of systems with
distant gas giants in nearly circular orbits may also be increased.

\section{Summary}

We have investigated orbital circularization 
due to planet growth through accreting disk gas.
We have analytically derived the differential equations 
for evolutions of orbital eccentricity $e$ and semimajor axis $a$
and numerically integrated them to discuss the solutions.

The motivation of these calculations is to examine 
our scenario for the formation of the distant gas giants
in nearly circular orbits, which are recently being discovered by
direct imaging surveys.
Our scenario is based on the conventional core accretion model
as follows: i) Icy cores accrete from planetesimals
in inner regions at $a \la 30$AU, ii) they are scattered outward by a nearby gas giant
to acquire highly eccentric orbits, iii) their orbits are circularized 
through accretion of local protoplanetary disk gas,
and iv) through the local gas accretion, the planets become gas giants.
We started our calculations after step ii).
For given initial $e$ and $a$, we followed the process in step iii).

For highly eccentric orbits, the planet spends 
most of time in the outer regions where disk gas has
higher specific orbital angular momentum than the planet.
Since the gas accretion rate from the disk is regulated by
envelope contraction except for final gas accretion phase,
we assume that disk gas accretion rate is constant
within one orbit.
Even in the final phase when the accretion rate is limited by
the supply of gas due to global disk accretion, the assumption
is valid if steady disk accretion is established. 
Thus, specific angular momentum of the planet increases
with planet accretion, resulting in circularization of the planetary orbit.
Energy dissipation by collision between disk gas and the planet
also induces the eccentricity damping.

Just after step ii), core's pericenter distance must be close to
it original location.
We found that pericenter distance is quickly raised by
the orbital circularization, so that perturbations of the gas giant
in the inner region can be neglected in the orbital circularization process. 
Thereby, we investigated orbital evolution of isolated planets
accreting disk gas.

The orbital evolutions that we found are:

\noindent
1) The eccentricity is reduced to $<0.2$ before
the planetary mass is increased by a factor of 10
(for example, if an icy core with $\sim 10M_{\oplus}$ 
starts gas accretion, its orbit is circularized 
with $e < 0.2$ before it acquires a Saturnian-mass.)

\noindent
2) The eccentricity damping is dominated over
the semimajor axis damping.
During $e$ is reduced from $\sim 1$ to zero,
$a$ is decreased only by a factor of 2.

\noindent
These show that planetary growth and 
orbital circularization concurrently proceed and
the orbital circularization is very efficient.
If we take into account the effect of bow shock for supersonic
incident gas flow, the orbital circularization becomes slower,
but it is still efficient enough to account for the observed
orbital properties of distant gas giants.
The orbit is left in large orbital radii, which
are about half of the semimajor axes that the scattered 
icy cores initially acquire. 

We performed the population synthesis calculation 
by incorporating the fitting formulas for the eccentricity and
semimajor axis damping by planet mass growth to indeed show
that the damping is efficient
and giants with $e \la 0.2$ are left in distant regions at $a \sim 30$--300AU.
However, with more detailed prescription using the formulas derived here, 
the fraction of systems
that have such distant jupiters is as small as $\sim 0.1\%$,
which is lower by a factor 4 than that predicted in \citet{ida13} using simpler
prescription.

We also consider the effect of the finite disk size.
If the eccentric orbits of the scattered cores
are deviated from the protoplanetary disk near their apocenters,
their semimajor axes shrink to a quarter of the disk sizes.
In other words, if observations show a concentration of
distant gas giants at some orbital radius,
it could reflect typical sizes of the protoplanetary disks,
in a similar way that the pile-up location of hot jupiters
could reflect the size of magnetospheric cavity (the size of the disk inner edge)
where type II migration could be stalled. 

We thank Prof. Andrew Youdin for valuable and useful comments as a referee.
We also thank Takayuki Muto and Taku Takeuchi for discussions.
Our study was supported by JSPS KAKENHI Grant Number 23103005.


\end{document}